\documentclass[a4paper,fleqn,usenatbib]{mnras}

\usepackage[T1]{fontenc}
\usepackage{ae,aecompl}

\usepackage{graphicx}   
\usepackage{amsmath}    
\usepackage{amssymb}    
\usepackage{graphics}
\usepackage{epsfig}
\usepackage{txfonts}
\usepackage{flushend}
\usepackage{blindtext}

\title[Highly-ionized metals in the vicinity of GRBs]{Highly-ionized metals as probes of the circumburst gas in the \\ natal regions of gamma-ray bursts}

\author[K.~E.~Heintz et al.]{K.~E.~Heintz$^{1,2,3}$,\thanks{E-mail: keh14@hi.is}
D.~Watson$^{2,3}$,
P.~Jakobsson$^{1}$,
J.~P.~U.~Fynbo$^{2,3}$,
J.~Bolmer$^{4}$,
\newauthor
M.~Arabsalmani$^{5,6}$,
Z.~Cano$^{7}$,
S.~Covino$^{8}$,
V.~D'Elia$^{9}$,
A.~Gomboc$^{10}$,
J.~Japelj$^{11}$,
\newauthor
L.~Kaper$^{12}$,
J.-K.~Krogager$^{13,3}$,
G.~Pugliese$^{11}$,
R. S\'{a}nchez-Ram\'{\i}rez$^{14}$,
J.~Selsing$^{3}$,
\newauthor
M.~Sparre$^{15}$,
N.~R.~Tanvir$^{16}$,
C.~C.~Th\"one$^{7}$,
A.~de~Ugarte~Postigo$^{7,3}$,
\& S.~D.~Vergani$^{6}$
\\
$^{1}$Centre for Astrophysics and Cosmology, Science Institute, University of Iceland, Dunhagi 5, 107 Reykjav\'ik, Iceland \\
$^{2}$Cosmic Dawn Center, Niels Bohr Institute, University of Copenhagen, Juliane Maries Vej 30, 2100 Copenhagen \O, Denmark \\
$^{3}$Dark Cosmology Centre, Niels Bohr Institute, University of Copenhagen, Juliane Maries Vej 30, 2100 Copenhagen \O, Denmark \\
$^{4}$European Southern Observatory, Alonso de C\'ordova 3107, Vitacura, Casilla 19001, Santiago 19, Chile \\
$^{5}$IRFU, CEA, Universit\'e Paris-Saclay, F-91191 Gif-sur-Yvette, France \\
$^{6}$GEPI, Observatoire de Paris, PSL Research University, CNRS, Place Jules Janssen, F-92190 Meudon, France \\
$^{7}$Instituto de Astrof\'isica de Andaluc\'ia (IAA-CSIC), Glorieta de la Astronom\'ia s/n, E-18008, Granada, Spain \\
$^{8}$INAF -- Osservatorio Astronomico di Brera, Via E. Bianchi 46, I-23807 Merate (LC), Italy \\
$^{9}$Space Science Data Center - Agenzia Spaziale Italiana, via del Politecnico, s.n.c., I-00133, Roma, Italy \\
$^{10}$Centre for Astrophysics and Cosmology, University of Nova Gorica, Vipavska 11c, 5270 Ajdov\v s\v cina, Slovenia \\
$^{11}$Anton Pannekoek Institute for Astronomy, University of Amsterdam, Science Park 904, 1098 XH Amsterdam, The Netherlands \\
$^{12}$Astronomical Institute Anton Pannekoek, University of Amsterdam, PO Box 94249, 1090 GE Amsterdam, the Netherlands \\
$^{13}$Institut d'Astrophysique de Paris, CNRS-UPMC, UMR7095, 98bis bd Arago, 75014 Paris, France \\
$^{14}$INAF, Istituto Astrofisica e Planetologia Spaziali, Via Fosso del
Cavaliere 100, I-00133 Roma, Italy \\
$^{15}$Institut f\"ur Physik und Astronomie, Universit\"at Potsdam, Karl-Liebknecht-Str.\,24/25, 14476 Golm, Germany \\
$^{16}$Department of Physics \& Astronomy and Leicester Institute of Space \& Earth Observation, University of Leicester, University Road, \\ Leicester LE1 7RH, United Kingdom \\
}

\date{Accepted 2018. Received 2018; in original form 2018}

\pubyear{2018}

\begin{document}
	\label{firstpage}
	\pagerange{\pageref{firstpage}--\pageref{lastpage}}
	\maketitle

\begin{abstract}
We present here a survey of high-ionization absorption lines in the
afterglow spectra of long-duration gamma-ray bursts (GRBs) obtained with the
VLT/X-shooter spectrograph. Our main goal is to investigate the circumburst
medium in the natal regions of GRBs. Our primary focus is on the
N\,\textsc{v}\,$\lambda\lambda$\,1238,1242 line transitions, but we
also discuss other high-ionization lines such as O\,\textsc{vi}, C\,\textsc{iv}
and Si\,\textsc{iv}. We find no correlation between the column density of
N\,\textsc{v} and the neutral gas properties such as metallicity, H\,\textsc{i}
column density and dust depletion, however the relative velocity of
N\,\textsc{v}, typically a blueshift with respect to the neutral gas, is found
to be correlated with the column density of H\,\textsc{i}. This may be
explained if the N\,\textsc{v} gas is part of an H\,\textsc{ii} region hosting
the GRB, where the region's expansion is confined by dense, neutral gas in the
GRB's host galaxy. We find tentative evidence (at $2\sigma$ significance) that
the X-ray derived column density, $N_{\mathrm{H,X}}$, may be correlated with
the column density of N\,\textsc{v}, which would indicate that both
measurements are sensitive to the column density of the gas located in the
vicinity of the GRB. We investigate the scenario where N\,\textsc{v} (and also
O\,\textsc{vi}) is produced by recombination after the corresponding atoms have
been stripped entirely of their electrons by the initial prompt emission, in
contrast to previous models where highly-ionized gas is produced by
photoionization from the GRB afterglow.
\end{abstract}




\section{Introduction}     
\label{sec:intro}

Cosmic lighthouses such as gamma-ray bursts (GRBs) allow the interstellar medium (ISM) of their host galaxies and foreground absorbers to be studied in great detail \citep[e.g.][]{Jakobsson04} up to high redshifts. In particular, they enable investigations of the ISM in terms of chemical abundances \citep{Fynbo06,Prochaska07,Cucchiara15}, dust content \citep{Savaglio03,Schady10,Zafar11,Zafar13}, kinematics \citep{Prochaska08a,Arabsalmani15,Arabsalmani18}, and the very local circumburst medium \citep{Moller02,Prochaska06,Prochaska08b,Fox08,CastroTirado10,Schady11}. The large majority of GRB afterglow spectra show damped Lyman-$\alpha$ (DLA) absorption, i.e., a neutral hydrogen column density of $N$(H\,\textsc{i}) > $2\times 10^{20}$ cm$^{-2}$ \citep[][and Tanvir et al., submitted]{Vreeswijk04,Jakobsson06,Fynbo09,Cucchiara15}, typically with higher H\,\textsc{i} column densities than seen in DLAs toward quasars \citep[e.g.][]{Noterdaeme09,SanchezRamirez16}. This is consistent with the scenario where the GRB explodes in the inner region, or in a dense, star-forming cloud, within its host galaxy \citep{Bloom02,Fruchter06,Prochaska07,Vreeswijk07,Vreeswijk11,Fynbo08,Svensson10,Lyman17}.

While the neutral gas-phase ISM in the host galaxy can be studied in great detail with GRBs, it is much more difficult to examine their immediate environments. Due to the hardness of the prompt emission and the extreme luminosity across all wavelengths, GRBs are expected to destroy dust and fully or at least partly ionize the circumburst gas, out to a few tens of pc \citep{Waxman00,Fruchter01,Perna03,Watson06,Watson13}. This can potentially erase any traces of the physical conditions of the circumburst material from their optical afterglow spectra. Highly ionized absorption features or X-ray observations of the GRBs might therefore be the only spectroscopically way to probe the circumburst medium. The observed soft X-ray absorption column density is typically larger than that of the neutral gas in GRB host galaxies \citep{Galama01,Watson07,Campana10,Schady11}, which could indicate that the X-ray absorption probes a significant column of hot, ionized metals that is transparent to UV photons. Another plausible scenario is that the He in the natal H\,\textsc{ii} region of the GRB is responsible for most of the soft X-ray absorption \citep{Watson13,Krongold13}. The absorption is likely due to the much larger population of He atoms, and since the abundance of He saturates the He-ionizing UV continuum out to about ten pc from the GRB, it can remain in a neutral or singly-ionized state. This scenario also supports previous models where dust is destroyed and metals are stripped of all their electrons by the GRB out to distances of tens to hundreds of pc \citep{Waxman00,Fruchter01,Perna03,Watson06,Watson13}.

Since the relative location of the absorbing gas cannot be resolved in X-ray spectroscopy, the observed column density probes the total gas column along the line of sight. For this reason the soft X-ray absorption might originate in the intergalactic medium (IGM) or in intervening absorbers. This was argued to be the cause for the observed correlation between the X-ray absorbing column density and redshift \citep{Behar11,Arcodia16}. However, \cite{Watson12} and \cite{Watson13} showed that this trend is largely a consequence of the higher gas column density in GRB hosts at high redshifts \citep[similar to what is observed for main-sequence galaxies, e.g.][]{Magdis12} and that at low-$z$, the dust bias (excluding high X-ray absorbing column density GRBs) is more severe than at high-$z$. Only at the highest redshifts ($z\gtrsim 4$) is it plausible that a significant component of absorption from intervening matter could be present \citep{Starling13}.

The long-standing problem of the origin of the soft X-ray absorption could therefore be solved by comparing the high-ionization absorption features seen in optical GRB afterglows to the X-ray derived column density. If a correlation is found between these two quantities, it may be possible to robustly relate the X-ray absorption to the immediate environment of the GRB. In this paper we compile a list of all high-redshift ($z\gtrsim 1.7$), long-duration GRBs with good quality X-shooter spectra to investigate the high-ionization absorption features in GRB afterglows. Such highly ionized absorption lines, in particular N\,\textsc{v}, has already been studied with high-resolution data though only in small samples \citep{Prochaska08b,Fox08}. Our sample is three times as large as those previously published, and by combining the individual samples we are now able to study the high-ionization absorption features with better statistics.

\begin{table*}
	\centering
	\begin{minipage}{0.75\textwidth}
		\caption{X-Shooter GRB sample and log of observations. GCN circulars that reported the gamma-ray burst detection and the subsequent X-shooter afterglow follow-up observations (if available). $^{a}$Time of spectroscopic follow-up after GRB trigger; $^{b}$Spectral resolution in the UVB and VIS arm. \textbf{References.} (1)~\citet{Markwardt09}; (2)~\citet{Malesani09a}; (3)~\citet{Bissaldi09}; (4)~\citet{Malesani09b}; (7)~\citet{Saxton11}; (8)~\citet{Wiersema11}; (9)~\citet{Beardmore12}; (10)~\citet{Sbarufatti12}; (11)~\citet{Kruhler12}; (12)~\citet{Pagani12a}; (13)~\citet{Malesani12}; (14)~\citet{Immler12}; (15)~\citet{Hartoog12}; (16)~\citet{Pagani12b}; (17)~\citet{Tanvir12b}; (18)~\citet{Lien13}; (19)~\citet{Hjorth13}; (20)~\citet{Ukwatta13}; (21)~\citet{Xu13}; (22)~\citet{Bissaldi14}; (23)~\citet{Xu14a}; (24)~\citet{DAvanzo14}; (25)~\citet{Xu14b}; (26)~\citet{Lien15}; (27)~\citet{Pugliese15}; (28)~\citet{Melandri15}; (29)~\citet{DeUgartePostigo15}; (30)~\citet{Ukwatta15}; (31)~\citet{Xu15}; (32)~\citet{DAvanzo16}; (33)~\citet{Pugliese16}; (34)~\citet{Mereghetti16}; (35)~\citet{Tanvir16}.}
		\centering
		\begin{tabular}{lccccccl}
			\noalign{\smallskip} \hline \hline \noalign{\smallskip}
			GRB & Trigger (UT) & $\Delta t^{a}$ & Mag$_{\mathrm{AB}}$ & Exp. time & ESO prog. ID & $\mathcal{R}_{\mathrm{UVB/VIS}}^{b}$ & Notes and refs. \\
			& (hh:mm:ss) & (hr) & (Acq.) & (s) &   & (km s$^{-1}$) &   \\
			\noalign {\smallskip} \hline \noalign{\smallskip}
			090809 & 17:31:14 & 10.2 & 21.0 & $4\times1800$ & 60.A-9427(A) & 53.6\,/\,31.3 & (1,2)   \\
			090926A & 21:55:48 & 22.0 & 17.9 & $4\times600$ & 60.A-9427(A) & 49.2\,/\,28.6 & (3,4)   \\
			100425A & 02:50:45 & 4.00 & 20.6 & $4\times 600$ & 85.A-0009(B) & 52.5\,/\,30.0 & (5,6) \\
			111008A & 22:12:58 & 8.50 & 21.0 & $4\times2400$ & 88.A-0051(B) & 46.2\,/\,26.5 & (7,8) \\
			120119A & 04:04:30 & 1.40 & 17.0 & $4\times 600$ & 88.A-0051(B) & 39.5\,/\,22.7 & (9) \\
			120327A & 02:55:16 & 2.10 & 18.8 & $4\times600$ & 88.A-0051(B) & 33.3\,/\,19.4 & (10,11)  \\
			120815A & 03:55:21 & 1.69 & 18.9 & $4\times600$ & 89.A-0067(B) & 44.1\,/\,25.4 & (12,13)   \\
			120909A & 01:42:03 & 1.70 & 21.0 & $2\times600$ & 89.A-0067(B) & 56.6\,/\,33.0 &   RRM, (14,15) \\
			121024A & 02:56:12 & 1.80 & 20.0 & $4\times600$ & 90.A-0088(B) & 41.7\,/\,24.0 & (16,17) \\
			130408A & 21:51:38 & 1.90 & 20.0 & $2\times600$ & 91.C-0934(C) & 39.5\,/\,22.7 & (18,19)  \\
			130606A & 21:04:39 & 7.10 & 19.0 & $7\times600$ & 91.C-0934(C) & 37.0\,/\,21.4 & (20,21) \\
			141028A & 10:54:46 & 15.4 & 20.0 & $4\times600$ & 94.A-0134(A) & 53.6\,/\,30.9 & (22,23)   \\
			141109A & 05:49:55 & 1.90 & 19.2 & $2\times1200$ & 94.A-0134(A) & 41.7\,/\,24.2 & (24,25)  \\
			150403A & 21:54:16 & 10.8 & 19.1 & $4\times600$ & 95.A-0045(A) & 44.8\,/\,26.1 & (26,27) \\
			151021A & 01:29:12 & 0.75 & 18.2 & $7\times600$ & 96.A-0079(B) &  50.0\,/\,29.1 & RRM, (28,29) \\
			151027B & 22:40:40 & 5.00 & 20.5 & $4\times600$  & 96.A-0079(A) & 55.6\,/\,32.3 & (30,31) \\
			160203A & 02:13:10 & 0.30 & 18.0 & $11\times600$ & 96.A-0079(A) & 37.5\,/\,21.7 & RRM, (32,33) \\
			161023A & 22:39:24 & 3.00 & 17.5 & $2\times600$ & 98.A-0055(A) & 46.2\,/\,26.5 & (34,35) \\
			\noalign{\smallskip} \hline \noalign{\smallskip}
		\end{tabular}
		\label{tab:obs}
	\end{minipage}
\end{table*}

The paper is structured as follows. In Section~\ref{sec:sel} we present our sample, discuss the imposed selection criteria and evaluate the neutral gas properties (such as $z_{\mathrm{GRB}}$, H\,\textsc{i} column density, metallicity, and dust depletion) of the GRBs in our sample. In Section~\ref{sec:results} we present our results, with a specific focus on N\,\textsc{v} which is the best constrained high-ionization absorption line. We provide a discussion of our data in Section~\ref{sec:disc}, where we also propose a new scenario for the origin of highly-ionized metals in the vicinity of GRBs. Finally, we conclude on our work in Section~\ref{sec:conc}.

Throughout the paper we assume a standard, flat cosmology with $H_0 = 67.8$\,km\,s$^{-1}$\,Mpc$^{-1}$, $\Omega_m = 0.308$ and $\Omega_{\Lambda}=0.692$ \citep{Planck16}. Abundances are expressed relative to solar, i.e. [X/H] = $\log N(\mathrm{X})/N(\mathrm{H}) - \log N(\mathrm{X})_{\odot}/N(\mathrm{H})_{\odot}$, using solar abundances from \cite{Asplund09}.

\section{The GRB sample}    \label{sec:sel}

The sample of GRB afterglows studied here was built by extracting all long-duration GRBs at $z\gtrsim 1.7$ from the VLT/X-shooter GRB (XS-GRB) legacy survey (Selsing et al., in preparation). Table~\ref{tab:obs} compiles the information of each of the GRBs in our study, including: The time of the observation, the spectroscopic sequence, observing programme, spectral resolution and references for the high-energy detection and X-shooter follow-up.
After examining each individual spectrum we imposed additional observational selection criteria to exclude bursts with poor S/N or insufficient coverage of the high-ionization lines (see Sect.~\ref{ssec:sel}). In Sect.~\ref{ssec:lowion} we report the low-ionization properties of the sample such as the systemic redshift, H\,\textsc{i} column density, gas-phase metallicity, and dust depletion.

\subsection{Observations and data reduction}\label{ssec:obs}

The majority of GRBs observed with the VLT/X-shooter spectrograph in our sample were detected by the Burst Alert Telescope (BAT) on-board the \textit{Niel Gehrels Swift Observatory} satellite \citep{Gehrels04}. The three exceptions are the GRBs\,090926A, 141028A, and 161023A, where the first two were detected by \textit{Fermi}-LAT/GBM \citep{Atwood09,Meegan09} onboard the \textit{Fermi} Gamma-Ray Space Telescope and the last was detected by the \textit{INTEGRAL} satellite \citep{Winkler03}.

The localization of each burst was distributed within a few minutes through the GCN which allowed for quick follow-up observations with VLT. Three of the 18 GRB afterglows in our sample (GRBs\,120909A, 151021A and 160203A) were observed with the rapid-response mode \citep[RRM;][]{Vreeswijk10} available at the ESO/VLT. All bursts were observed within 24 hours after trigger (see also below). 

The VLT/X-shooter spectrograph covers the spectral wavelength range from 3000\,\AA~-- 24800\,\AA, divided into three spectroscopic arms \citep{Vernet11}. The UVB arm covers 3000\,\AA~-- 5500\,\AA, the VIS arm 5500\,\AA~-- 10200\,\AA~and the NIR arm 10200\,\AA~-- 24800\,\AA. In the large majority of cases, the spectra were obtained following an ABBA nodding pattern, the slits were aligned with the parallactic angle, selecting slit widths of $1\farcs0$, $0\farcs9$ and $0\farcs9$ for the three arms, respectively. For this setup the nominal instrumental resolutions in the UVB, VIS and NIR arms are 70 km\,s$^{-1}$, 40 km\,s$^{-1}$ and 55 km\,s$^{-1}$. If the seeing on the night of observation is smaller than the slit width projected on the sky, the delivered resolution is enhanced. In the VIS and NIR arms the resolution is determined by fitting a range of telluric absorption lines and extracting the full-width-at-half-maximum (FWHM). Since no tellurics fall in the UVB arm from which we can measure the resolution directly, we extrapolated the ratio of the observed-to-nominal resolution in the VIS arm and assume the same ratio for the observed-to-nominal resolution in the UVB arm. For each burst we list the delivered resolution in the UVB and VIS arm in Table~\ref{tab:obs}.

All the science spectra were reduced in a consistent manner for the XS-GRB legacy sample. The exact procedure is described in Selsing et al. (in prep), so in the following we will only briefly summarize the data reduction strategy. First, the raw spectra were processed through a cosmic-ray removal algorithm \citep{VanDokkum01}. Then, each of the exposures were separately reduced using the VLT/X-shooter pipeline, version \texttt{2.7.1} or newer \citep{Goldoni06,Modigliani10}, and were subsequently combined in the post-processing. The pipeline produces a flat-fielded, rectified and wavelength-calibrated 2D spectrum for every frame in the UVB, VIS and NIR arm. To extract the final 1D spectra from the combined 2D frames we model the trace using a running Voigt-profile fit. All wavelengths reported throughout the paper are in vacuum and are shifted to the heliocentric velocity. Finally, the spectra were corrected for the Galactic extinction in the line-of-sight using the dust maps of \cite{Schlegel98} from the most recent update \citep{Schlafly11}.

\begin{table*}
	\begin{minipage}{0.75\textwidth}
\caption{Low ionization properties and N\,\textsc{v} detection summary. \textbf{References.} (1)~\citet{Cucchiara15}; (2)~\citet{Delia10}; (3)~\citet{Sparre14}; (4)~\citet{Wiseman17}; (5)~\citet{Delia14}; (6)~\citet{Kruehler13}; (7)~\citet{Friis15}; (8)~\citet{Hartoog15}; (9)~Tanvir et al. (submitted); (10)~Th\"one et al. (in preparation); (11)~Pugliese et al. (in preparation); (12)~de Ugarte Postigo et al. (submitted).}
		\centering
		\begin{tabular}{lcccccccl}
			\noalign{\smallskip} \hline \hline \noalign{\smallskip}
			GRB & $z_{\mathrm{GRB}}$ & $\log N(\textsc{H\,i})$ & [X/H] & [X/Fe] & Ion & $\log N$(N\,\textsc{v}) & $\delta v$ & Refs. \\
			& & (cm$^{-2}$) & && & (cm$^{-2}$) & (km s$^{-1}$) &  \\
			\noalign {\smallskip} \hline \noalign{\smallskip}
			090809 & 2.73706 & $21.70\pm 0.20$ & $-0.57\pm 0.10$ & $0.30\pm 0.31$ & Si & $<14.43$ & $\cdots$ & (1)  \\
			090926A & 2.10694  & $21.60\pm 0.07$  & $-1.85\pm 0.10$ & $0.34\pm 0.13$ & S & $14.30\pm 0.19$ & $4\pm 2$ & (2) \\
			100425A & 1.75640 & $21.05\pm 0.10$ & $-0.96\pm 0.42$ & $\cdots$ & Fe & $<14.63$ & $\cdots$ & (1) \\ 
			111008A & 4.99146 & $22.30\pm 0.06$  & $-1.70\pm 0.10$ & $0.05\pm 0.13$ & S & $>14.00$ & $8\pm 10$ & (3) \\
			120119A & 1.72883 & $22.70\pm 0.20$ & $-0.79\pm 0.42$ & $0.36\pm 0.28$ & Si & $<15.15$ & $\cdots$ & (4) \\
			120327A & 2.81482 & $22.01\pm 0.09$  & $-1.17\pm 0.11$ & $0.56\pm 0.15$ & Zn & $13.56\pm 0.03$ & $0\pm 1$ & (5) \\
			120815A &  2.35820 & $21.95\pm 0.10$  & $-1.15\pm 0.12$ & $1.01\pm 0.10$ & Zn & $14.60\pm 0.18$ & $-6\pm 7$ & (6) \\
			120909A & 3.92882 & $21.70\pm 0.10$  &  $-0.66\pm 0.11$ & $0.88\pm 0.16$ & S & $14.75\pm 0.11$ & $-14\pm 4$  & (1) \\
			121024A & 2.30244 &  $21.88\pm 0.10$  & $-0.40\pm 0.12$ & $0.85\pm 0.04$ & Zn & $<14.35$ & $\cdots$  & (7) \\
			130408A & 3.75792 & $21.80\pm 0.10$  & $-1.24\pm 0.12$ & $0.11\pm 0.26$ & S & $14.44\pm 0.07$ & $-16\pm 2$  & (1) \\
			130606A & 5.91278 & $19.91\pm 0.02$  & $-1.30\pm 0.08$ & $0.79\pm 0.13$ & Si & $14.50\pm 0.04$ & $-83\pm 1$  & (8) \\
			141028A & 2.33327 & $20.60\pm 0.15$  & $-0.50\pm 0.38$ & $0.57\pm 0.21$ & Si & $14.28\pm 0.10$ & $-104\pm 20$  & (4) \\
			141109A & 2.99438 & $22.10\pm 0.20$  & $-1.40\pm 0.22$ & $0.84\pm 0.37$ & Zn & $\gtrsim14.85$ & $13\pm 15$  & (9,10) \\
			150403A & 2.05707 & $21.80\pm 0.20$ & $-0.80\pm 0.35$ & $1.31\pm 0.43$ & S & $14.73\pm 0.14$ & $-32\pm 3$  & (9,10) \\
			151021A & 2.32975 & $22.20\pm 0.20$ & $-1.11\pm 0.20$ & $1.15\pm 0.28$ & Si & $14.80\pm 0.46$ & $34\pm 5$  & (9,10) \\
			151027B & 4.06469 & $20.50\pm 0.20$ & $-1.62\pm 0.24$ & $0.34\pm 0.28$ & Si & $<13.98$ & $\cdots$  & (9,10) \\
			160203A & 3.51871 & $21.75\pm 0.10$ & $-1.26\pm 0.11$  & $0.59\pm 0.25$ & S & $<13.58$ & $\cdots$  & (9,11) \\
			161023A & 2.71067 & $20.96\pm 0.05$ & $-1.24\pm 0.09$ & $0.45\pm 0.11$ & Si & $13.66\pm 0.08$ & $-114\pm 6$  & (12) \\
			\noalign{\smallskip} \hline \noalign{\smallskip}
		\end{tabular}
		\label{tab:lowion}
	\end{minipage}
\end{table*}

\subsection{Selection criteria} \label{ssec:sel}

After examining the individual quality of all the spectra we imposed additional observational and brightness constraints to remove bursts with poor signal-to-noise (S/N) ratios (effectively eliminating the majority of sources for which the mean $S/N\lesssim 3$ per wavelength element). The criteria we imposed for the GRB follow-up observations are: \textit{i}) that the GRB was observed with X-shooter within 24 hours after the GRB trigger, \textit{ii}) the brightness measured from the X-shooter acquisition image is brighter than 21.0 mag and \textit{iii}) there is spectral coverage of the Ly$\alpha$ profile and at least the N\,\textsc{v} doublet of the high-ionization transitions, effectively requiring that the galaxies hosting the bursts are located at $z\gtrsim 1.7$. In total, 18 GRBs observed with the VLT/X-shooter fulfill these criteria, see Table~\ref{tab:obs} (for the observational details) and Sect.~\ref{ssec:lowion} and Table~\ref{tab:lowion} (for the low-ionization properties of each). 

The N\,\textsc{v} transition is of particular interest to us since it has been shown that the majority of this highly ionized gas must be located within approximately 1 kpc of the GRB explosion site \citep{Prochaska06} or perhaps even within $\lesssim 10$ pc \citep{Prochaska08b}. Studying these events, together with the low-ionization properties of the host galaxy, allows us to constrain the physical properties of their circumburst medium. Moreover, this line is located in the red wing of the absorption trough of the Ly$\alpha$ line and is easily detected compared to the majority of the high-ionization species that are located in the Ly$\alpha$ forest. The final sample of GRBs observed with VLT/X-shooter presented here is three times the size of previous surveys for N\,\textsc{v} absorption in GRB afterglows \citep{Prochaska08b,Fox08} and extends the redshift range out to $z\sim 6$. These former two surveys found a high detection rate of N\,\textsc{v} in 6/7 GRB afterglows, however, five bursts were common to both samples. From our sample here we can independently assess the detection rate.

For completeness, we also survey the high-ionization species of N\,\textsc{v}$\,\lambda\,1238,1242$, O\,\textsc{vi}$\,\lambda\,1031,1037$,  C\,\textsc{iv}$\,\lambda\,1548,1550$, Si\,\textsc{iv}$\,\lambda\,1393,1402$, S\,\textsc{iv}$\,\lambda\,1062$ and S\,\textsc{vi}$\,\lambda\,933,944$. The ionization potential to create these ions are 77.5, 113.9, 47.9, 33.5, 34.8 and 72.6 eV (see also Table~\ref{tab:161023a}). In addition to these, we also look for P\,\textsc{v}$\,\lambda\,1117,1128$, which to our knowledge have not been observed in GRB afterglow spectra before. Furthermore, we include absorption lines that trace the neutral gas, typically Si\,\textsc{ii} or Fe\,\textsc{ii} to 1) determine the systemic redshifts, $z_{\mathrm{GRB}}$, and 2) to measure the gas-phase metallicity of the galaxy hosting the GRB (see below). In the Appendix we provide notes on all the individual GRBs in our sample and the detected high-ionization lines. 

\subsection{Low-ionization properties} \label{ssec:lowion}

For each burst, the systemic redshift, $z_{\mathrm{GRB}}$, is defined by the position of the peak optical depth in the low-ionization lines (typically Fe\,\textsc{ii} or Si\,\textsc{ii}). We searched for high-ionization absorption features within $z_{\mathrm{GRB}} \pm 500$ km\,s$^{-1}$, representing the low-velocity components. We did not look for high-velocity components ($500 - 5000$ km\,s$^{-1}$ relative to $z_{\mathrm{GRB}}$), that likely trace outflows that can be driven to high velocities by the burst progenitor \citep[see e.g.,][]{Fox08,CastroTirado10}, since we are only interested in the immediate environment around the GRB. We list the measured $z_{\mathrm{GRB}}$ for each burst in the second column in Table~\ref{tab:lowion}. In the third column, the neutral hydrogen (H\,\textsc{i}) column density is reported for each GRB where we adopt the values reported in the single papers that examined the individual bursts (see references in the table) or those from the extensive compilation of GRB afterglow H\,\textsc{i} column densities by Tanvir et al. (submitted, and references therein). 

Except for GRBs\,141109A, 150403A, 151021A, 151027B, 160203A, and 161023A (de Ugarte Postigo et al., submitted), we adopt the gas-phase metallicities reported in the literature (see the references in Table~\ref{tab:lowion}), listed as [X/H], where the ion X is one of the volatile species, Si, S or Zn. In all cases, the dust depletion of the ion X relative to iron, [X/Fe], is also reported. For the remaining six bursts in our sample we use our own fitting tool to derive column densities and gas-phase metallicities of the low-ionization absorption features by performing a series of Voigt-profile fits to the absorption line profiles (the full set of absorption line fits to the XS-GRB sample will be published in Th\"one et al., in preparation). 

\begin{figure}
	\centering
	\includegraphics[width=9cm]{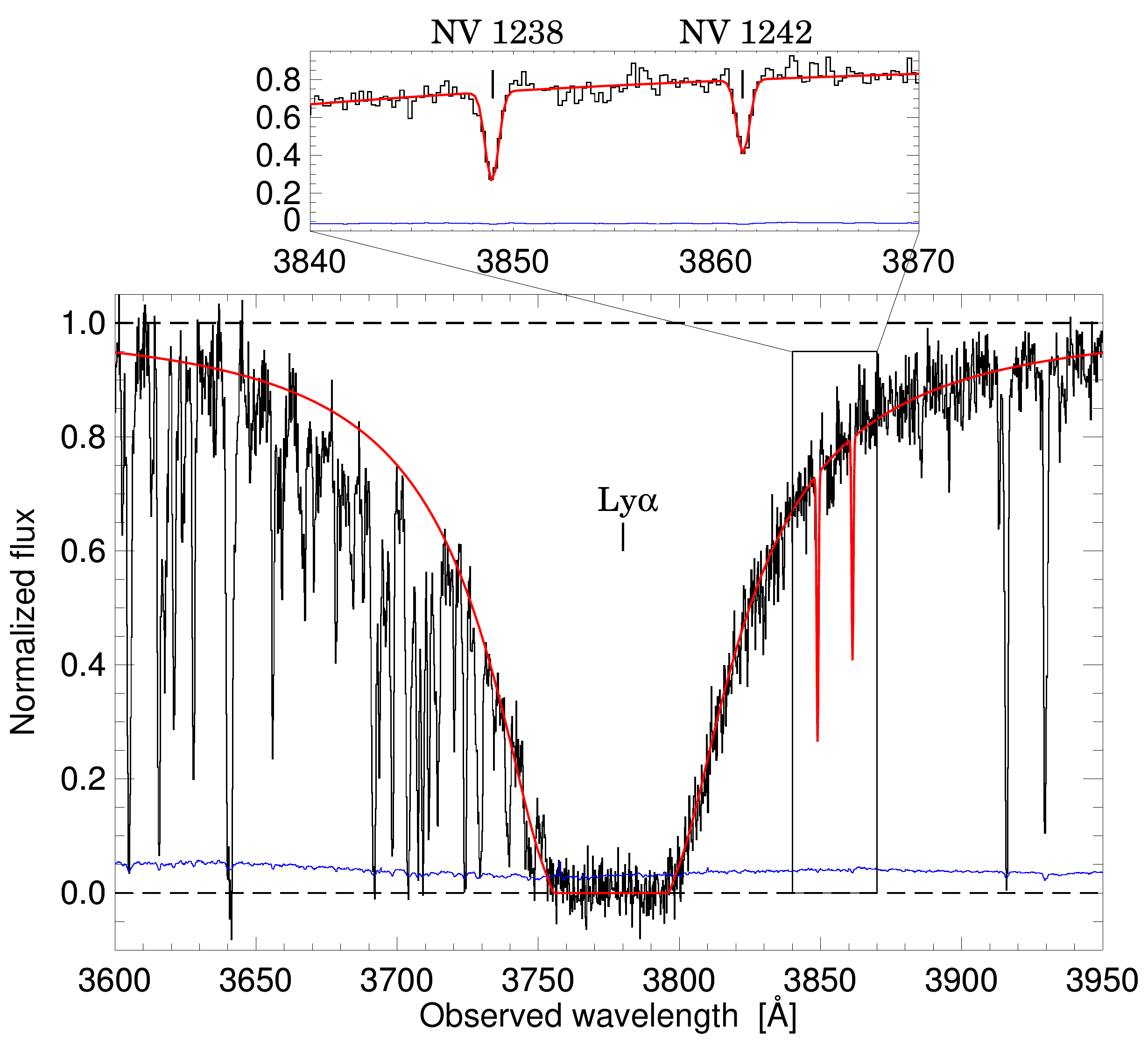}
	\caption{VLT/X-shooter spectrum of GRB\,090926A. In the bottom panel, the black solid line shows the normalized UVB arm spectrum centred on Ly$\alpha$ at $z\approx2.107$ with the associated error spectrum shown in blue. The best-fit Voigt profiles to the Ly$\alpha$ line and the N\,\textsc{v}\,$\lambda\lambda$\,1238,1242 doublet is shown as the solid red line. In the upper panel we show a zoomed-in region of the N\,\textsc{v} transitions and the same combined Voigt profile fit. 
	}
	\label{fig:090926_hinv}
\end{figure}

The code is developed in \texttt{Python} and is described in detail in \cite{Krogager18}. Simply, the software takes as input the instrumental resolution, the zero-point redshift and the H\,\textsc{i} column density and returns the redshift of the different velocity components ($z$), line width ($b$), and column density ($\log N$) for each species. The derived metallicities and depletion patterns are summarized in Table~\ref{tab:lowion} and are reported in detail for each burst in the Appendix. We follow the same procedure to derive the column densities of the high-ionization absorption features by fitting the various components and ionization species in each GRB afterglow spectrum independently. Since the N\,\textsc{v}\,$\lambda\lambda$\,1238,1242 doublet is located in the red wing of the Ly$\alpha$ absorption trough in bursts with large H\,\textsc{i} column densities ($\log N$(H\,\textsc{i}) $\gtrsim 21$), the continuum flux in this region will be underestimated. To measure the column density of N\,\textsc{v} we therefore fit the N\,\textsc{v} and Ly$\alpha$ lines simultanously, using a prior estimate of the H\,\textsc{i} column density from the already published values. This approach is shown in Fig.~\ref{fig:090926_hinv} for GRB\,090926A (with an H\,\textsc{i} column density of $\log N$(H\,\textsc{i}) $\gtrsim 21.60$), where it is also clearly demonstrated how the Ly$\alpha$ absorption affects the continuum level in the region around the N\,\textsc{v} doublet. For the non-detections of N\,\textsc{v}, 3$\sigma$ upper limits were calculated using $N$(N\,\textsc{v}) = $1.13\times 10^{20}\,W_{\mathrm{lim}}/\lambda_0^2 f$ where $W_{\mathrm{lim}}$ is the 3$\sigma$ upper limit on the rest-frame equivalent width in \AA~and $N$(N\,\textsc{v}) is in cm$^{-2}$.

\section{Results}    \label{sec:results}

\begin{table}
	\caption{Voigt profile fits for GRB\,161023A at $z=2.71067$. The ionization potential (IP) for each transition is also listed.}
	\begin{tabular}{cccccc}
		\noalign{\smallskip} \hline \hline \noalign{\smallskip}
		Ion & IP & $v_0$ & $b$ & $\log N$ & $\log N_{\mathrm{total}}$  \\
		& eV & (km s$^{-1}$)  & (km s$^{-1}$) & ($N$ in cm$^{-2}$) & ($N$ in cm$^{-2}$) \\ 
		\noalign{\smallskip} \hline \noalign{\smallskip}
		N\,\textsc{v} & 77.5 & $-114\pm 6$ & $30\pm 10$ & $13.66\pm 0.08$ & $13.66\pm 0.08$ \\
		O\,\textsc{vi} & 113.9 & $-118\pm 6$ & $78\pm 9$ & $14.59\pm 0.04$ & $14.59\pm 0.04$  \\
		C\,\textsc{iv} & 47.9 & $-130\pm 2$ & $44\pm 3$ & $14.91\pm 0.07$ & $14.93\pm 0.07$ \\
		&& $-8\pm 4$ & $22\pm 7$ & $13.60\pm 0.04$ &  \\ 
		Si\,\textsc{iv} & 33.5 & $-117\pm 4$ & $26\pm 7$ & $14.47\pm 0.18$ & $14.50\pm 0.17$ \\
		&& $-27\pm 13$ & $44\pm 17$ & $13.30\pm 0.13$ &  \\ 
		&& $-168\pm 65$ & $47\pm 41$ & $13.56\pm 0.75$ &  \\ 
		S\,\textsc{iv} & 34.8 & $-104\pm 8$ & $58\pm 12$ & $14.66\pm 0.07$ & $14.66\pm 0.07$ \\
		S\,\textsc{vi} & 72.6 & $-121\pm 6$ & $52\pm 8$ & $14.17\pm 0.06$ & $14.17\pm 0.06$ \\
		P\,\textsc{v} & 65.0 & $-39\pm 6$ & $49\pm 9$ & $13.54\pm 0.06$ & $13.54\pm 0.06$ \\
		\noalign{\smallskip} \hline \noalign{\smallskip}
	\end{tabular}	
	\label{tab:161023a}
\end{table}

\begin{figure}
	\centering
	\includegraphics[width=8.8cm,page=1]{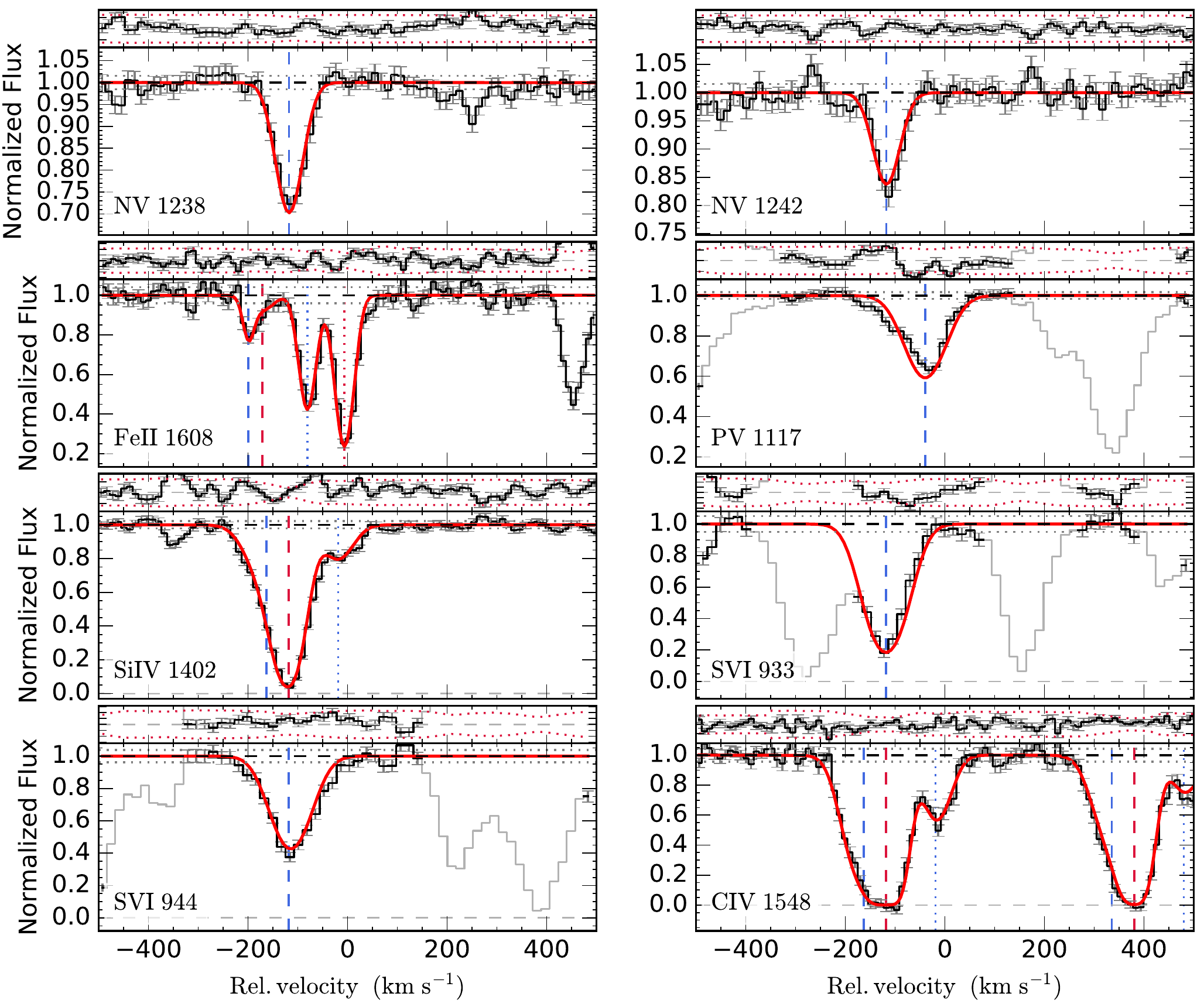}
	\qquad
	\includegraphics[width=8.9cm,page=2]{GRB161023A_nhion.pdf}
	\caption{GRB161023A: Normalized high-ionization absorption-line profiles on a velocity scale relative to $z_{\mathrm{GRB}}$. Voigt profile fits are shown in red (solid line) with the center of each velocity component marked with dotted or dashed lines. The part of the spectrum shown in grey is masked out in the fit. The residual spectrum to the best fit is shown at the top of each panel.}
	\label{fig:161023A_hion}
\end{figure}

The results of the Voigt profile fits for each burst are shown in the Appendix (Tables~\ref{tab:090809} to \ref{tab:160203a} and Figs.~\ref{fig:090809_hion} to \ref{fig:160203A_hion}) together with notes on all the GRB afterglows in our sample. We report the best-fit Voigt profile parameters in individual tables and show the corresponding fit in the figures belonging to each burst. In each of the figures, an example of one of the low-ionization absorption lines is included, where the peak optical depth of this line represents $z_{\mathrm{GRB}}$. In Table~\ref{tab:161023a} we list the results of the Voigt profile fits and show the corresponding best-fit combined line profiles in velocity-space in Fig.~\ref{fig:161023A_hion} for one of the bursts, GRB\,161023A, as an example (but see also de Ugarte Postigo et al., submitted).

In the majority of cases, the high-ionization species O\,\textsc{vi}, S\,\textsc{iv}, S\,\textsc{vi}, and P\,\textsc{v} are blended with the Ly$\alpha$ forest. Si\,\textsc{iv} and C\,\textsc{iv} are detected in all of the GRB afterglow spectra but are both heavily saturated. In this section we therefore focus mainly on the N\,\textsc{v} doublet that has the advantage of not being saturated (in most cases) compared to some of the other high-ionization lines, and is located in the red wing of the Ly$\alpha$ absorption trough. The medium resolution of the X-shooter spectrograph also ensures that the N\,\textsc{v} doublet is well resolved. Due to the varying quality of the spectra and since the Ly$\alpha$ absorption line and N\,\textsc{v} doublet lies in the bluemost part of the UVB arm for redshifts $z\approx 1.7$, the detection limit varies from approximately $\log N$(N\,\textsc{v}) = 13.5 to 15.2. The results of the N\,\textsc{v} column densities and 3$\sigma$ upper limits are summarized in Table~\ref{tab:lowion}, where we also list the relative velocity to $z_{\mathrm{GRB}}$. For the GRBs with N\,\textsc{v} detections we find average and median values of $\log N$(N\,\textsc{v})$_{\mathrm{avg}} = 14.36$ and $\log N$(N\,\textsc{v})$_{\mathrm{med}} = 14.50$, respectively. We obtain robust fits to all the N\,\textsc{v} lines, and we find that blending or line saturation is not an issue in the majority of cases (although see the individual tables for the few exceptions). As mentioned above, the column densities for N\,\textsc{v} are fitted simultaneously with the Ly$\alpha$ line and those values are reported in Table~\ref{tab:lowion} and in the individual tables in the Appendix. The measured column densities using this approach are, however, mostly consistent (within the $1\sigma$ errors) to the values obtained by fitting only the N\,\textsc{v} lines.

As the large ionization potential of N\,\textsc{iv} (as well as O\,\textsc{v} and S\,\textsc{v}) well exceeds the integrated galactic ionizing radiation from O and B stars that drops strongly above 54 eV due to the He\,\textsc{ii} ionization edge \citep{Bregman86}, an additional intense radiation source is required to produce N\,\textsc{v}. Furthermore, \cite{Prochaska08b} demonstrated that GRB afterglows can photoionize nitrogen up to N\,\textsc{v} if the gas is closer than $\approx 10$ pc. This, together with the fact that the relative velocities of the N\,\textsc{v} lines in general are coincident with the UV-pumped fine-structure lines of Si\,\textsc{ii} and Fe\,\textsc{ii} \citep{Prochaska08b}, implies that the observed N\,\textsc{v} absorption features are related to the material close to the GRB or its progenitor. 
The following sections are dedicated to investigate the relation between the characteristics of the N\,\textsc{v} gas and the physical properties of the GRB host galaxy, such as GRB redshift, ISM gas-phase metallicity, dust and metal column density and internal kinematics. 


\begin{figure}
	\centering
	\epsfig{file=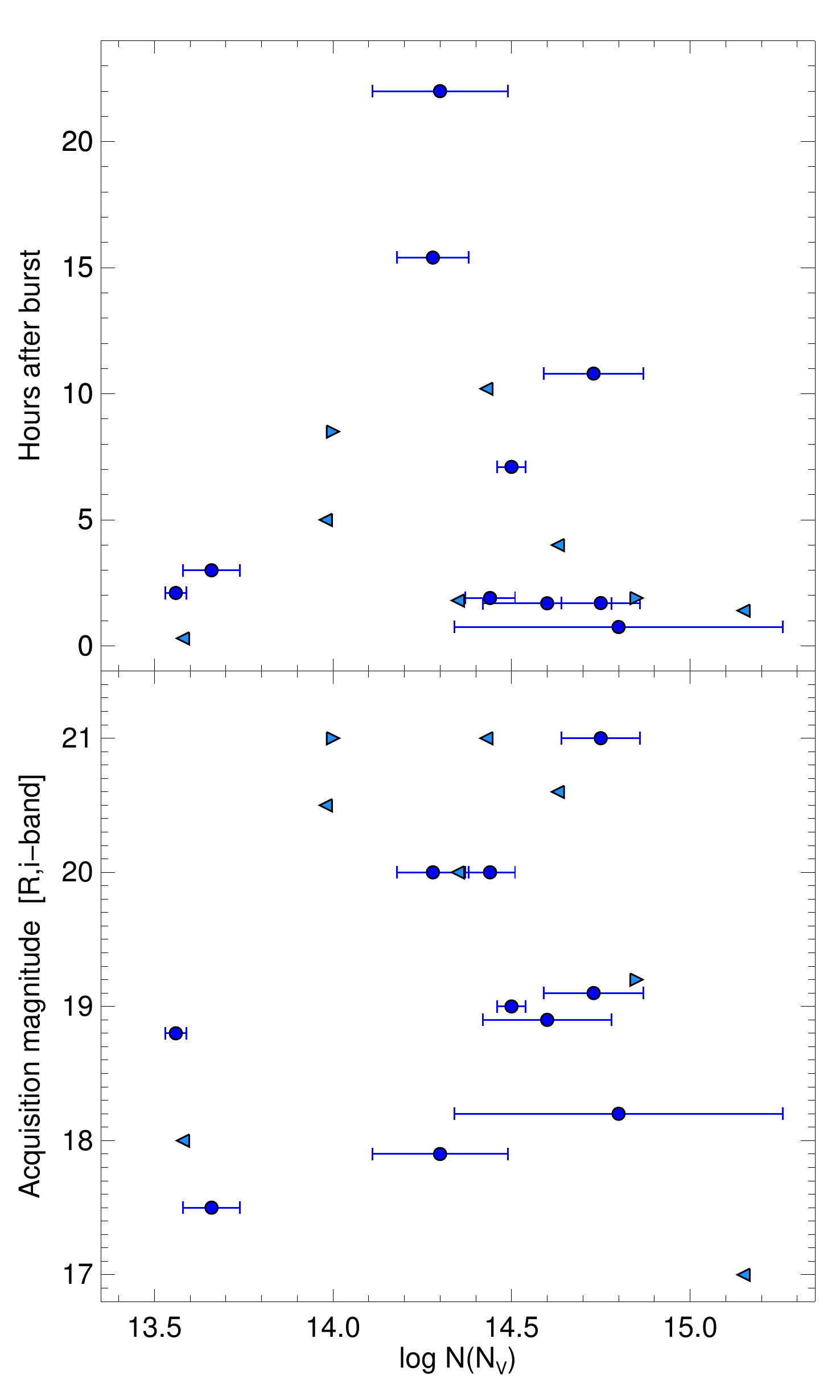,width=9cm}
	\caption{Detection of N\,\textsc{v} as a function of time of observation post-burst (upper panel) and acquisition magnitude of the afterglow (bottom panel). Blue dots denote bursts where N\,\textsc{v} is detected and a column density measured. Light blue triangles are non-detections and the corresponding 3$\sigma$ upper limit is instead shown.}
	\label{fig:nvdtmag}
\end{figure}

\begin{figure}
	\centering
	\epsfig{file=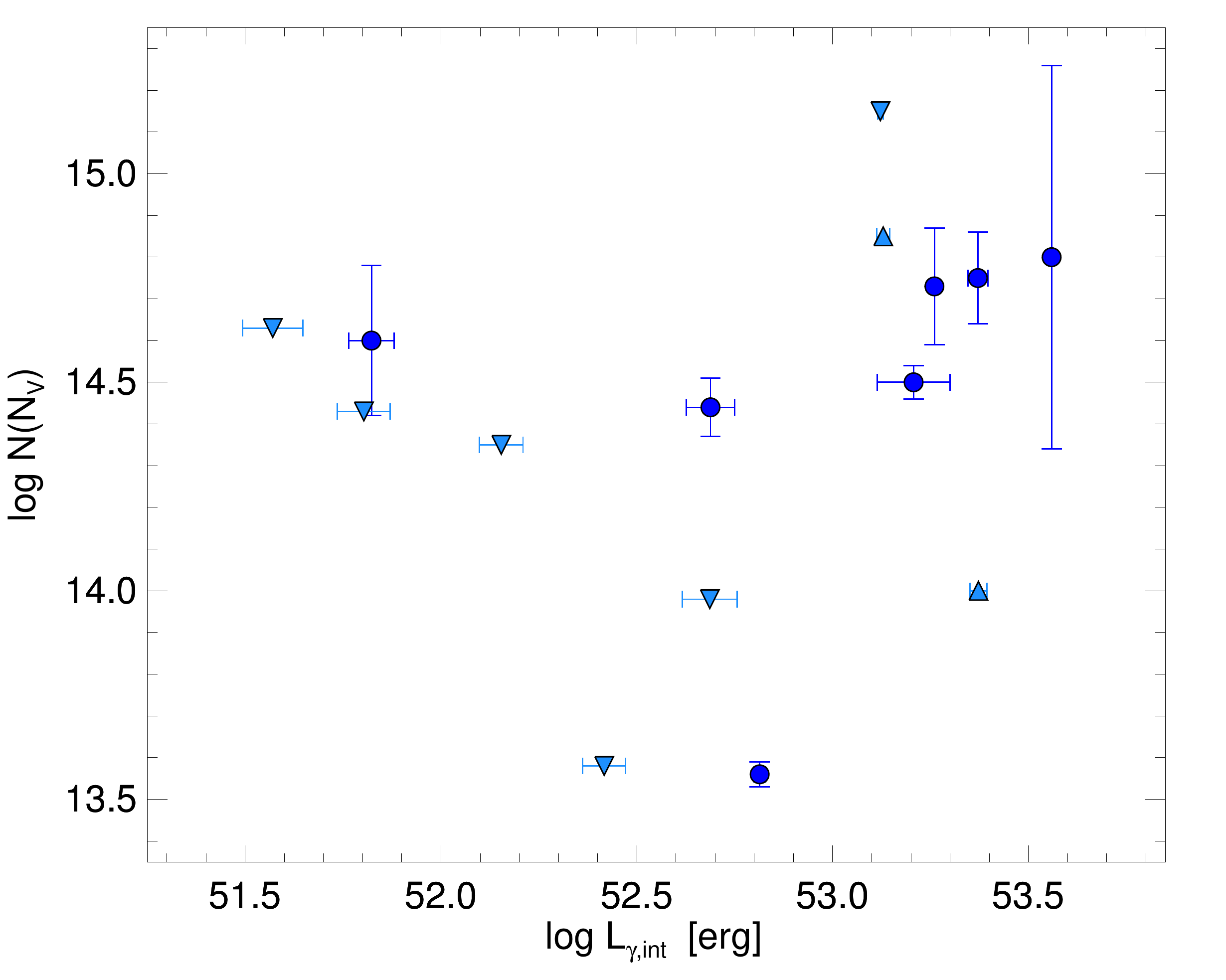,width=9cm}
	\caption{Detection of N\,\textsc{v} as a function of the intrinsic gamma-ray luminosity, $L_{\gamma,\mathrm{int}} = F_{\gamma,\mathrm{obs}}~4\pi~d_L^2~(1+z)^{-1}$, where $F_{\gamma,\mathrm{obs}}$ is the observed BAT fluence in the 15 -- 150 keV band. The full sample is shown except for GRBs\,090926A, 141028A and 161023A since these were not detected by \textit{Swift}/BAT. The same symbols are used as in Fig.~\ref{fig:nvdtmag}.}
	\label{fig:nvgrflu}
\end{figure}

\subsection{Detection rates}

We detect at least two of the surveyed high-ionization absorption features within a few hundred km\,s$^{-1}$ of $z_{\mathrm{GRB}}$ in all 18 GRB spectra. The detection rates are N\,\textsc{v} (seen in 12/18 cases), O\,\textsc{vi} (3/18), C\,\textsc{iv} (18/18), Si\,\textsc{iv} (18/18), S\,\textsc{iv} (5/18), S\,\textsc{vi} (6/18), and P\,\textsc{v} (4/18). The detection rates of the high-ionization lines that originate in the Ly$\alpha$ forest are to be treated as lower limits since they are, in the majority of cases, heavily blended or located in a spectral region with a poor S/N. Compared to earlier studies of high-ionization lines in GRB afterglows \citep[e.g.][]{Prochaska08b,Fox08}, the detection rate of N\,\textsc{v} is lower in our sample. Both of the previous samples found that N\,\textsc{v} was detected in 6/7 GRB afterglow spectra. The single burst, GRB\,060607, with a non-detection of N\,\textsc{v} down to $\log N$(N\,\textsc{v}) < 12.61, common to both samples, is unique in the sense that it has a very low H\,\textsc{i} column density. The physical properties of the line of sight to this GRB is therefore most likely significantly different.

While our sample show six non-detections of N\,\textsc{v}, the limits in the individual GRB afterglow spectra are almost at the same order as the derived N\,\textsc{v} column densities. To understand why N\,\textsc{v} is only observed in 12/18 GRB afterglow spectra, our first approach is to examine whether the detection rate is caused by observational constraints. In Fig.~\ref{fig:nvdtmag} we show the derived column density of N\,\textsc{v} as a function of hours post-burst at the time of observation and the measured acqusition magnitude (values taken from Selsing et al., in preparation). There appears to be no correlation between any of these parameters. We note, however, that four out of the six bursts with non-detections of N\,\textsc{v} were faint in the acquisition images at the time of observation, which would suggest that the cause of the non-detections are of observational origin.

Another scenario could be that the luminosity of the afterglow itself is not intense enough to produce highly-ionized metals. To test this scenario we examine the gamma-ray luminosity, $L_{\gamma,\mathrm{int}}$ (erg s$^{-1}$), of each burst and compare it to the N\,\textsc{v} column density. We extract the gamma-ray fluence, $F_{\gamma,\mathrm{obs}}$ (erg cm$^{-2}$), from the \textit{Swift}/BAT database in the observer frame energy band 15-150 keV. To convert this value to an intrinsic property we calculate $L_{\gamma,\mathrm{int}} = F_{\gamma,\mathrm{obs}}~4\pi~d_L^2~(1+z)^{-1}$ \citep[similar to][]{Lien16}, where $d_L$ is the luminosity distance. We caution that this is a simple approximation, and that a $k$-correction in principle also should be applied to account for the spectral shape. The results are shown in Fig.~\ref{fig:nvgrflu}. It is clear that the non-detections all belong to intrinsically faint GRBs (as already noted above from the acquisition images) except for GRB\,120119A for which we are, however, only able to put a poorly constrained upper limit on N\,\textsc{v}. For reference, GRB\,060607 (with a non-detection reported for N\,\textsc{v}) has an intrinsic gamma-ray fluence of $\log L_{\gamma,\mathrm{int}} = 52.71$. This value is around the median for the GRBs in our sample, suggesting that the non-detection in this case is likely due to the different nature of the galaxy hosting this GRB. For comparison we detect two cases with one and two orders of magnitudes higher N\,\textsc{v} column densities ($\log N$(N\,\textsc{v}) = $13.5 - 14.5$) with the same intrinsic gamma-ray luminosity, $\log L_{\gamma,\mathrm{int}} \approx 52.7$, as GRB\,060607. 

\begin{figure}
	\centering
	\epsfig{file=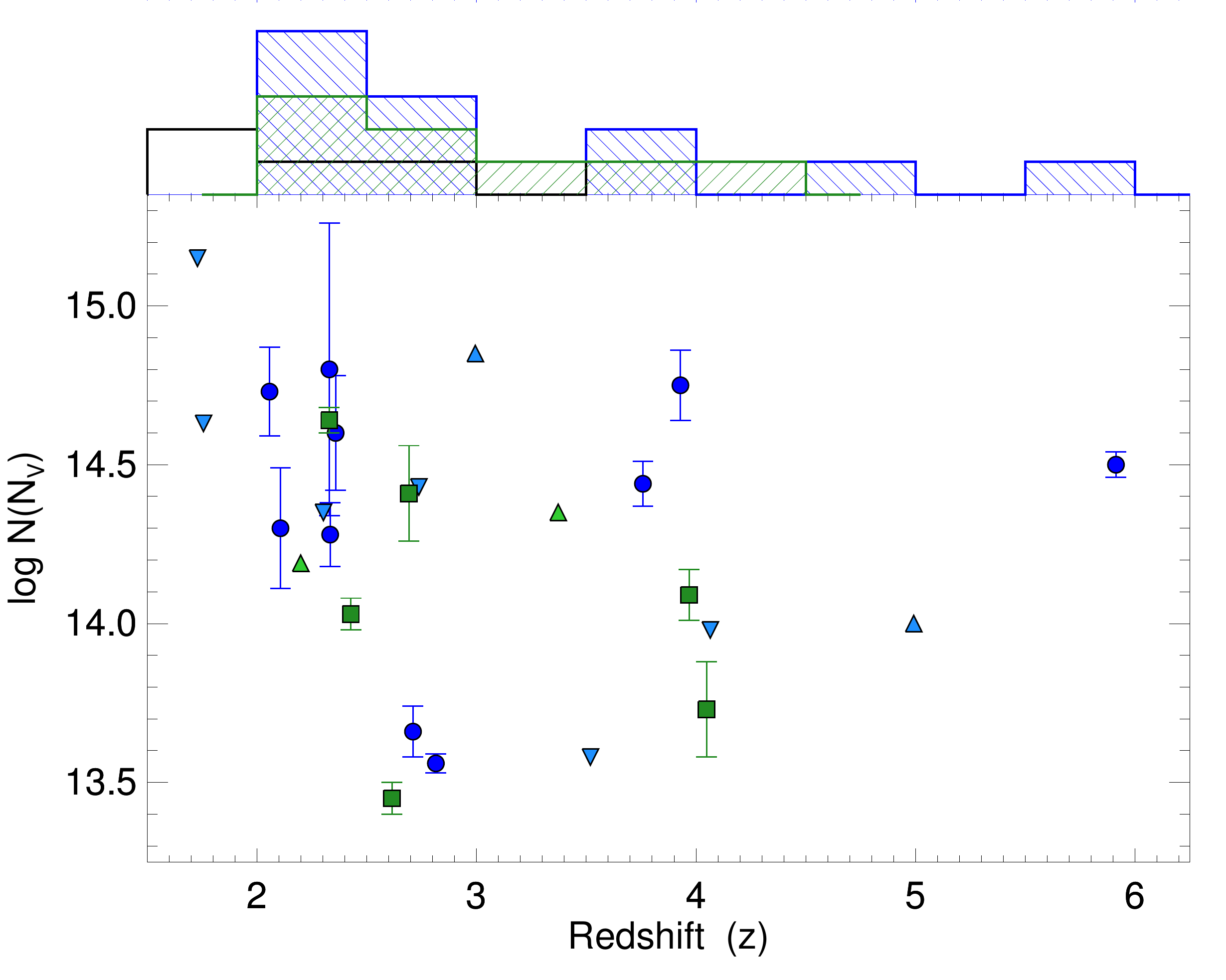,width=9cm}
	\caption{The column density of N\,\textsc{v} as a function of redshift. In the bottom panel, the same symbols are again used as in Fig.~\ref{fig:nvdtmag} and \ref{fig:nvgrflu} for GRBs in our sample. We have included N\,\textsc{v} detections from two previous GRB surveys \citep{Prochaska08b,Fox08}, shown here as the green squares and upward facing triangles (lower limits). In the upper panel we show the redshift distribution of the GRBs in our sample with (blue filled) and without (solid line) N\,\textsc{v} detections as histograms. The full sample spans a large redshift range of $1.7 < z < 6$. In addition, the distribution of the included GRBs with N\,\textsc{v} detections from previous surveys is shown as the green filled histogram.}
	\label{fig:nvz}
\end{figure}

\begin{figure*}
	\centering
	\epsfig{file=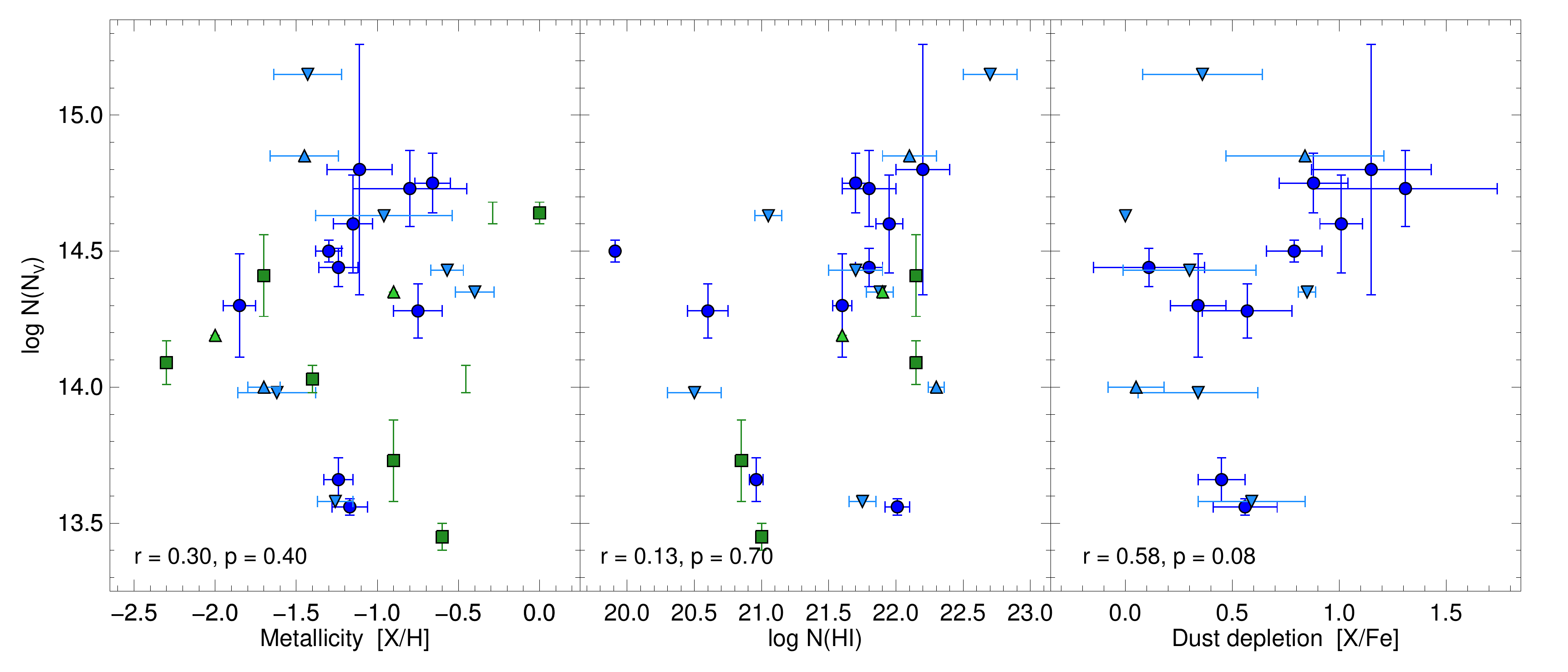,width=18cm}
	\caption{The column density of N\,\textsc{v} compared to low-ionization properties. We show the detection of N\,\textsc{v} in our sample, and the detections from the previous two GRB surveys for N\,\textsc{v}, as a function of metallicity (left panel), H\,\textsc{i} column density (middle panel), and dust depletion (right panel). In each panel the computed $r$ and $p$ values are shown. The same symbols are used as in Fig.~\ref{fig:nvz}. }
	\label{fig:nvlowion}
\end{figure*}

Since the total isotropic gamma-ray energy is correlated with the X-ray and optical afterglow brightness \citep{Nysewander09}, it is reasonable to expect that the production of highly ionized metals should be correlated with the radiation field from the GRB itself and the following afterglow radiation. The fact that the strength of N\,\textsc{v} depends on the intrinsic gamma-ray energy of the GRB then supports a scenario where N\,\textsc{v} traces the circumburst medium of the GRB. We will return to this issue in Sect.~\ref{ssec:cbm} where we provide further evidence that N\,\textsc{v} is a good probe of the metals in the molecular cloud or H\,\textsc{ii} region surrounding the GRB.

\subsection{Dependence of N\,\textsc{v} on neutral gas properties}

To explore the different factors that might influence the production of N\,\textsc{v}, we plot the measured N\,\textsc{v} column densities as a function of redshift in Fig.~\ref{fig:nvz} and as a function of metallicity, H\,\textsc{i} column density and dust depletion in Fig.~\ref{fig:nvlowion}. We include the GRBs from the two previous surveys for N\,\textsc{v} and high-ionization absorption \citep{Prochaska08b,Fox08} in both figures. Our sample spans a redshift range of $1.7 < z < 6$ and N\,\textsc{v} column densities of $13.5 \lesssim \log N$(N\,\textsc{v}) $ \lesssim 15.0$ (not including non-detections). 
The probability and strength of N\,\textsc{v} detection is not dependent on redshift, but we note that the two GRBs at $z>4$ have higher than average N\,\textsc{v} column densities. 

Comparing the probability and strength of the N\,\textsc{v} detection to the physical properties related to the neutral gas of the GRB hosts reveals that N\,\textsc{v} is not dependent on the H\,\textsc{i} column density and metallicity (see the computed linear Pearson correlation coefficients in each of the panels in the figure), consistent with previous conclusions. We note, however, that there is a tentative correlation with dust depletion at the $1\sigma$ level and that strong N\,\textsc{v} absorption is predominantly seen at high values of $N$(H\,\textsc{i}). Our sample spans metallicites of -1.9 < [X/H] < -0.4, neutral hydrogen column densities of 19.9 < $\log N$(H\,\textsc{i}) < 22.7 and dust depletion ratios of 0.1 < [X/Fe] < 1.3. We can therefore provide further evidence that the highly-ionized gas in GRBs is not directly related to the properties of the neutral ISM. This is in contrast to what is observed for QSO-DLAs with N\,\textsc{v} detections, where the detection probability is found to be dependent on metallicity \citep{Fox09}. Absorbing galaxies toward quasars are not expected to contain intense non-stellar radiation (like GRBs), so the high-ionization absorption in quasar DLAs likely originates in the ISM or in the halo gas \citep[which also explains the lower mean value of $\log N$(N\,\textsc{v}) = 13.78, and the large percentage ($\approx 87\%$) of non-detections][]{Fox09}.

\subsection{Velocity offset of  N\,\textsc{v}}

A crucial piece of information about the circumburst medium can be obtained if we can localize the GRB in the host galaxy and derive its immediate physical conditions.
However, deep, resolved imaging of GRB host galaxies is required to accurately locate the GRB explosition sites and to investigate the physical properties of their surrounding medium. Such maps are very challenging to obtain for GRB hosts at high redshifts \citep[][although see e.g. \cite{McGuire16} where \textit{HST} imaging was sufficient to resolve the location of the GRB\,130606A close to the centre of its host]{Tanvir12a}. Instead, we can examine the relative velocity of the main component of high-ionization absorption, assuming that it represents discrete clouds in the host galaxy, to see how it coincides with the low-ionization components tracing the neutral ISM gas. Detailed studies of the velocity components of high-ionization lines can also potentially reveal the scales of which the intervening gas absorbs the photons from the GRB afterglow, shown e.g. for GRB\,021004 \citep{Moller02,CastroTirado10}.

\begin{figure}
	\centering
	\epsfig{file=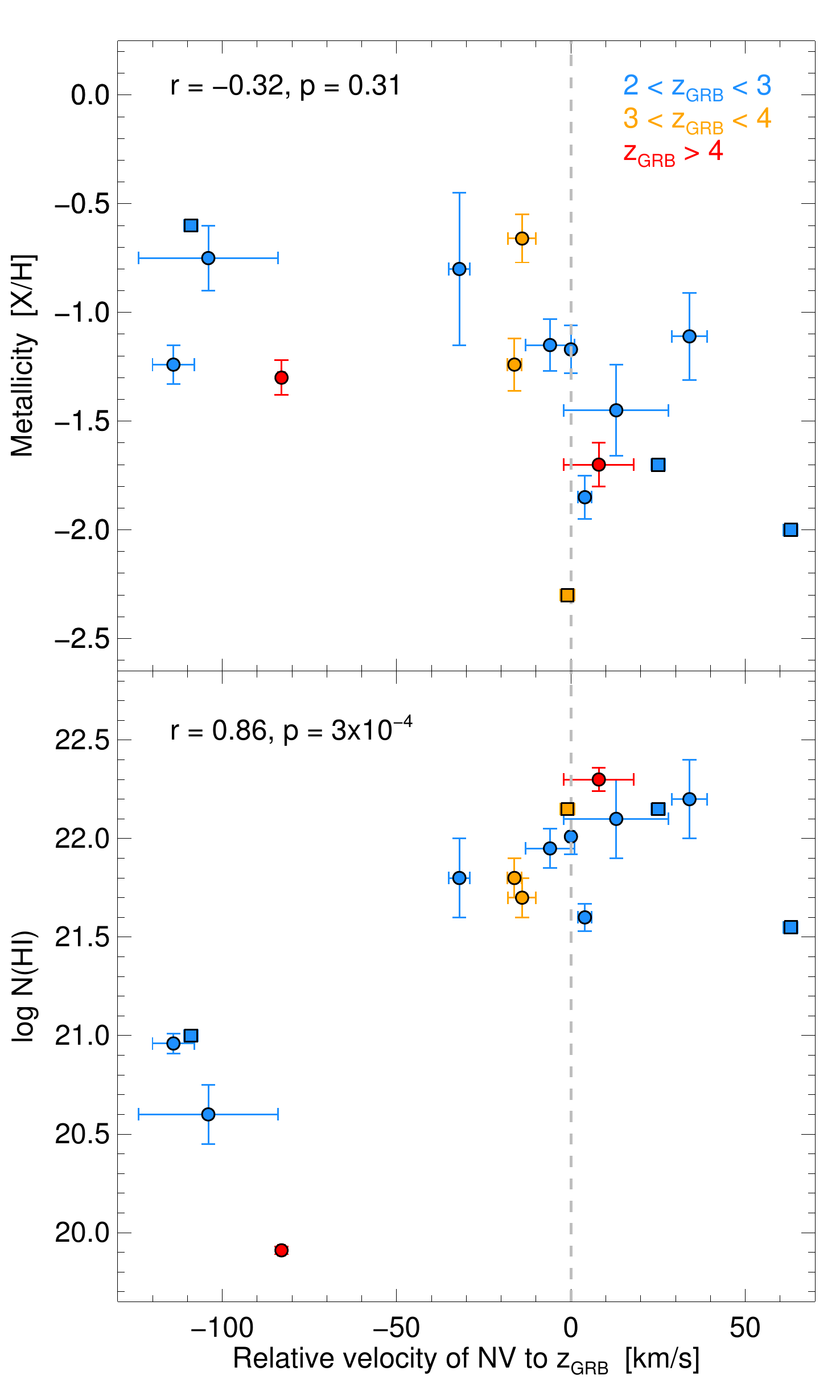,width=9cm}
	\caption{The relative velocity of the strongest component of N\,\textsc{v} compared to $z_{\mathrm{GRB}}$ as a function metallicity (upper panel) and H\,\textsc{i} column density (lower panel). The dots (our sample) and squares (bursts from the previous samples) representing the GRBs are color-coded as a function of redshift. In both panels the computed $r$ and $p$ values are shown. The dashed line intercept the GRBs for which the position of the strongest high-ionization absorption component is consistent with the strongest low-ionization absorption component in relative velocity space.}
	\label{fig:metvrel}
\end{figure}

We compare the relative velocity of the strongest N\,\textsc{v} component to $z_{\mathrm{GRB}}$ as a function of metallicity and H\,\textsc{i} column density in Fig.~\ref{fig:metvrel}. If the strongest, high-ionization components are aligned with the peak optical depth of the low-ionization profiles, they likely originate in the same cloud. If GRBs preferably occur in metal-poor environments it is reasonable to expect that the metal-rich GRB host galaxies would show high-ionization components separated from the strongest low-ionization metal components. We do not find any statistically significant correlation between the relative velocity of the high-ionization components to $z_{\mathrm{GRB}}$ as a function of metallicity. We do note, however, that at higher metallicities the scatter around $z_{\mathrm{GRB}}$ increases. This is likely due to a mass-metallicity scaling relation, where the more metal-rich GRBs are consistently more massive \citep[and thus physically larger;][]{Ledoux06,Arabsalmani18}. This allows for a greater range of internal velocities of possible star-forming clouds. 

We find that the relative velocity of N\,\textsc{v} to $z_{\mathrm{GRB}}$ depends on the H\,\textsc{i} column density of the GRB. While an approximate linear function could describe the relation between the two quantities quite well (see the $r$ and $p$ values in the bottom panel of Fig.~\ref{fig:metvrel}), an alternative is that the data instead show two populations: One with large H\,\textsc{i} column densities ($\log N$(H\,\textsc{i}) > 21.5) for which N\,\textsc{v} is aligned with $z_{\mathrm{GRB}}$ and one with low H\,\textsc{i} column densities ($\log N$(H\,\textsc{i}) < 21.0) for which the N\,\textsc{v} component is separated by $\approx 80 - 120$\,km\,s$^{-1}$ relative to the peak optical depth of the low-ionization profiles. While we do not find any correlation between N\,\textsc{v} and H\,\textsc{i} column densities, such that the immediate environment of the GRB is directly related to the H\,\textsc{i} column density, we argue that the data instead shows a confinement effect. We propose a scenario where the pressure of dense ($\log N$(H\,\textsc{i}) > 21.5) clouds confine the H\,\textsc{ii} region or molecular cloud, in which the GRB exploded, to the part of the galaxy with the largest amount of neutral gas. In more diffuse clouds ($\log N$(H\,\textsc{i}) < 21.0) the star-forming region is not as confined by the gas pressure. From the lower panel of Fig.~\ref{fig:metvrel} it can also be seen that only blueshifts of high-ionization components are measured with a relative velocity of $\approx -100$\,km\,s$^{-1}$. This would also be explained by the confinement scenario where the diffuse H\,\textsc{i} gas is not dense enough to counterbalance the effect of stellar winds originating from massive stars in the H\,\textsc{ii} region. Stellar winds causing the expansion of the H\,\textsc{ii} region is also of the order $\approx 100$\,km\,s$^{-1}$ \citep{Dyson77}, which would explain why we only observe the blueshifted component: the absorbing material is expanding and thus move towards us in the line of sight to the GRB explosion site. If this scenario is true, then we would also expect that the relative velocity of the fine-structure lines are slightly less blueshifted compared to N\,\textsc{v}, since the are typically measured to be at distances around a few hundred pc from the GRB. Only one of the GRBs in all three high-ionization line samples have a simultaneous measurement of N\,\textsc{v} (at $\delta v \approx -80$ km\,s$^{-1}$) and a fine-structure line \citep[GRB\,130606A][]{Hartoog15}, which shows a slightly smaller blueshift of Si\,\textsc{ii}* at $\delta v \approx -50$ km\,s$^{-1}$. For more of such cases, the relative velocity of the fine-structure and high-ionization would be valuable to examine. 

\begin{figure}
	\centering
	\epsfig{file=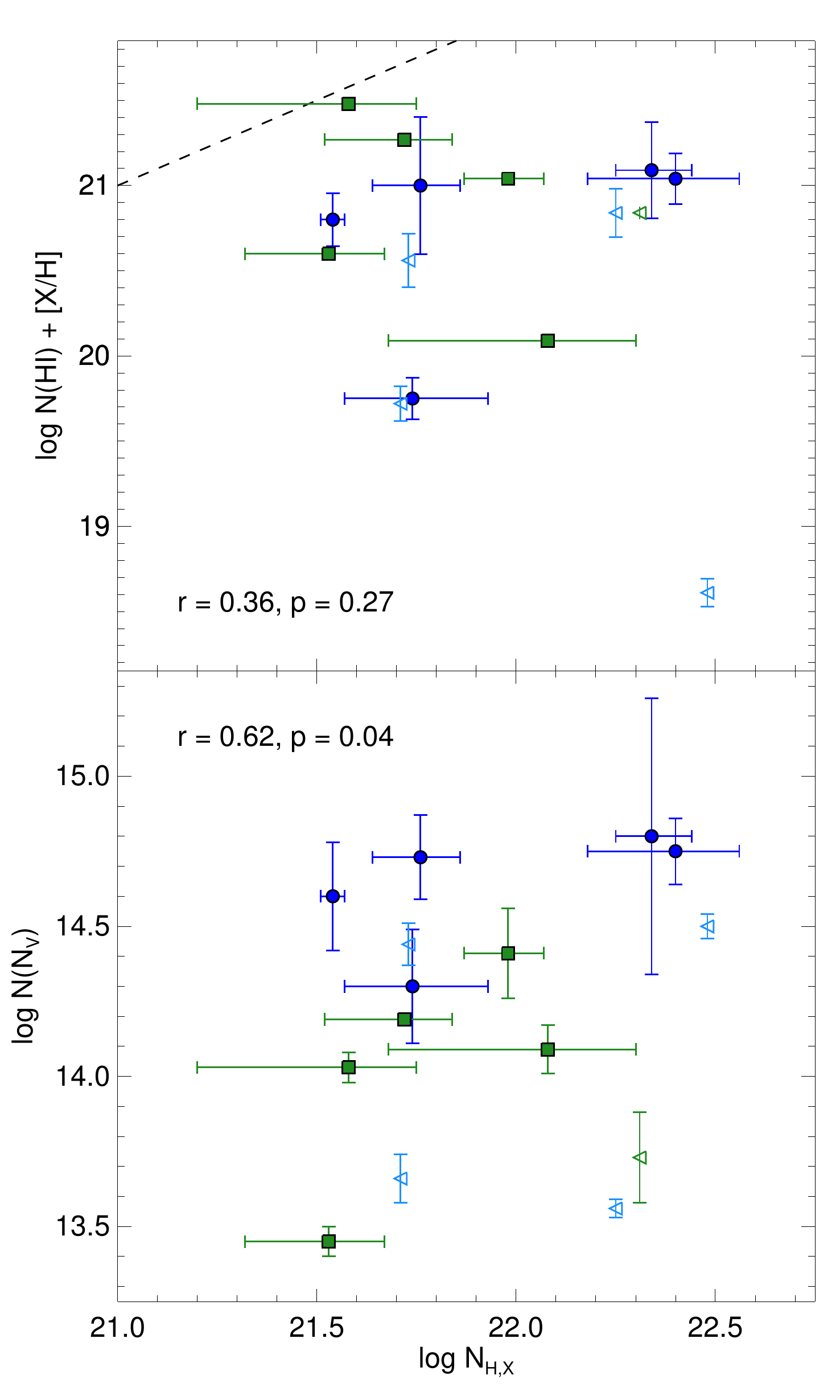,width=9cm}
	\caption{The \textit{Swift}/XRT soft X-ray derived metal column density, $N_{\mathrm{H,X}}$, as a function of the UV/optical metal column density, $\log N$(H\,\textsc{i}) + [X/H], (upper panel) and the N\,\textsc{v} column density (bottom panel). In both panels the computed $r$ and $p$ values are shown. In the upper panel the 1:1 ratio between the two distinct metal column densities is represented as the dashed line. Only the GRBs from our sample and the two previous samples with N\,\textsc{v} detections are plotted in both panels. Open triangles denote GRBs with upper limits on $N_{\mathrm{H,X}}$.}
	\label{fig:nvxmetcol}
\end{figure}

\subsection{The circumburst gas} \label{ssec:cbm}

The majority of GRB afterglows observed with \textit{Swift}/XRT show soft X-ray absorption in excess of Galactic \citep[e.g.][]{Galama01,Watson02}. While the X-ray absorption is believed to be dominated by metals, its origin has yet to be fully understood. Some authors claim that it traces the IGM along the line of sight \citep[e.g.][]{Behar11,Campana15} due to the increasing soft X-ray column density with redshift, however, no such trend has been observed e.g. toward quasars \citep{Eitain13}. Since the Galactic X-ray absorption is also well accounted for \citep{Willingale13}, this indicates that the observed excess X-ray derived absorption column density is likely located at the GRB redshift. Furthermore, the derived X-ray absorption is typically found to be larger than the column density of neutral hydrogen \citep{Watson07,Watson13,Campana10} and the metal column density of the neutral gas \citep{Galama01,Schady10,Zafar11,Greiner11} related to the dust extinction, $A_V$, of the ISM \citep{Zafar13}. This is also demonstrated in the upper panel of Fig.~\ref{fig:nvxmetcol} where we compare the the X-ray derived column density and the optically derived metal column density of the neutral ISM ($\log N$(H\,\textsc{i}) + [X/H]) for the bursts with N\,\textsc{v} detections from our sample and the two previous samples of \cite{Prochaska08b} and \cite{Fox08}. The X-ray column densities are either adapted from the literature \citep[e.g.,][but see the notes in the Appendix]{Campana10} or from the \textit{Swift}/XRT repository\footnote{\url{http://www.swift.ac.uk/xrt_spectra/}}. We confirm the trend of significantly larger X-ray derived metal column densities and observe no correlation with the optically derived metal column density. This indicates that the soft X-ray absorption is not dominated by the neutral gas of the ISM in GRB host galaxies.

This suggests that the large X-ray absorption originates in a much smaller region, possibly close to the GRB. The main difficulty in studying the circumburst medium of GRBs is that they are so energetic that they are expected to destroy dust and ionize gas to very large distances from the burst, thereby largely removing this circumburst gas from tracers in the optical afterglow spectra. Previously, it has been postulated that the majority of the X-ray absorption might originate from highly ionized gas \citep{Schady11} or from helium \citep{Watson13} in the immediate, circumburst environment of the GRB, constituting the H\,\textsc{ii} region. Therefore, the most viable tool to study the circumburst material is likely from hot gas signatures or to observe the gas at X-ray wavelengths. No strong correlations are yet observed, however, thereby constraining the use of the soft X-ray absorption as a diagnostic tool to study the circumburst medium of GRBs. 

\cite{Prochaska08b} suggests that both the N\,\textsc{v} column density and the X-ray absorption should be sensitive to the metals in the vicinity of the GRB. If N\,\textsc{v} is produced by photoionization from the GRB afterglow, and the X-ray photons are absorbed at the same approximate distance, we would expect a correlation between the two measurements. In the lower panel of Fig.~\ref{fig:nvxmetcol} we demonstrate for the first time that there is indeed evidence for such a correlation. Combining our sample with the two previous samples we find that by computing the linear Pearson correlation coefficients for the N\,\textsc{v} column density as a function of the X-ray derived absorption column density yield $r = 0.62$ and $p=0.04$ (excluding upper limits), i.e., the null hypothesis is rejected at more than $2\sigma$. GRB\,141028A is located outside the plotting region with an upper limit on $\log N_{\mathrm{H,X}}$ of <23 cm$^{-2}$ to better illustrate the linear correlation. Subtracting the optically derived metal column density from the X-ray absorption column yields a similar result, indicating that the metals in the neutral gas-phase in the ISM provide a negligible contribution to the X-ray absorption. The fact that there is marginal evidence for a correlation between N\,\textsc{v} and the X-ray absorption also suggests that the majority of the X-ray derived column density originates in the host galaxies of GRBs and not e.g., in the intergalactic medium (IGM) or intervening systems. The scenario where the X-ray absorption traces the H\,\textsc{ii} region or (former) molecular cloud in which the GRB exploded is also supported by the observed correlation between the X-ray column density and the surface luminosity \citep{Lyman17} since both are expected to be related to a region of intense star formation. 

\section{Discussion} \label{sec:disc}

\subsection{Highly-ionized gas from recombination}

While it has previously been shown by \citet{Prochaska08b} that N\,\textsc{v} can be produced by photoionization from the GRB afterglow, their model relies on no prior ionization of the circumburst gas, e.g.\ by the initial prompt emission from the GRB. Based on the correlation between N\,\textsc{v} and the X-ray derived metal column density, $N_{\mathrm{H,X}}$, we infer that the relative abundance of nitrogen is related to the total metal column density inferred from X-ray absorption. In the following we examine whether N\,\textsc{v} absorption could be dominated by the recombination of nitrogen stripped of electrons by the prompt emission \citep[as suggested by e.g.][]{Watson13} out to a few pc. We demonstrate that the observed column density of N\,\textsc{v} (and also O\,\textsc{vi} for a few cases) can be reproduced by the recombination of nitrogen (and oxygen) atoms using simple approximations. However, the temporal evolution of the column densities of these ions should be quite pronounced.

The recombination rate for a given transition is $R=\alpha\,n_e\,n_i$, where $\alpha$ is the recombination rate for the specific atom. The recombination time is therefore of the order $t_{\mathrm{rec}}\propto 1/(\alpha\,n_e)$. For the gas temperature ($T\approx 10^4$ K) and density ($n_e = 4\times10^3$ cm$^{-3}$) in a typical H\,\textsc{ii} region, and assuming that the total recombination coefficient $\alpha = 3\times 10^{-9}$ for N\,\textsc{v} \citep[though this value is two orders of magnitude higher than the sum of the radiative and dielectronic recombination rate coefficients]{Goshu17}, this yields a recombination time of $t_{\mathrm{rec}} \approx 10^{5}$\,s (i.e.\ about a day).
Then, assuming the relative solar abundance of N to H is $6.7\times 10^{-5}$ \citep{Asplund09}, we find that an absorber with $10^{22}$\,cm$^{-2}$ (typically observed in X-rays) yields $6.7\times 10^{17}$ nitrogen atoms cm$^{-2}$.
Assuming all the nitrogen of interest is completely stripped of electrons immediately following the burst, and every other source of N\,\textsc{iv} ionizing photons is negligible, the fraction of nitrogen in the N\,\textsc{v} state can be determined following the recombination series and assuming the electron density remains approximately constant over this time. Under the order of magnitude assumption that the recombination coefficients of N\,\textsc{v}--N\,\textsc{viii} are the same, the general equation for the expected column density of N\,\textsc{v} can be written as
$ N(\mathrm{N\,\textsc{v}}) =
\frac{1}{6}N(\mathrm{N}) \times
\left( \alpha\,n_e\,\Delta t_{\mathrm{rest}} \right)^3
e^{- \alpha\,n_e\,\Delta t_{\mathrm{rest}}}$,
where $N(\mathrm{N}) = N_{\mathrm{H,X}}\times(\mathrm{N/H})_{\odot}$, e.g.\ $N(\mathrm{N}) = 6.7\times 10^{17}$\,cm$^{-2}$ for an X-ray absorber with $10^{22}$\,cm$^{-2}$.
At the typical time of observation, a somewhat high density of $\sim2\times10^4$\,cm$^{-3}$ is required so that $\sim10^{-3}$ of the nitrogen atoms will have recombined from N\,\textsc{viii} to N\,\textsc{v} to give a value of $\log N$(N\,\textsc{v})/cm$^{-2} \sim 14.6$, consistent with the linear correlation between $N_{\mathrm{H,X}}$ and N\,\textsc{v} found in Sect.~\ref{ssec:cbm}. In the above we have assumed the recombination rate from \cite{Goshu17} for N\,\textsc{v} and that this is the same for N\,\textsc{v} to N\,\textsc{viii}. If the recombination rate is an order of magnitude lower or higher this would then yield electron densities of $2\times10^5$\,cm$^{-3}$ and $2\times10^3$\,cm$^{-3}$ instead, respectively.

Since O\,\textsc{vi} is expected to be produced by recombination in this scenario as well, we would expect the relative abundance of N\,\textsc{v} to O\,\textsc{vi} to be consistent with the solar ratio. This must be the case since the recombination rate is roughly the same for both ions and since N\,\textsc{v} and O\,\textsc{vi} are both lithium-like atoms with three electrons. In the recombination scenario, both atoms are therefore populated on the same time scales from fully ionized to N\,\textsc{v} and O\,\textsc{vi}, respectively. From the five bursts with robust measurements of both N\,\textsc{v} and O\,\textsc{vi} in our sample, and from \cite{Fox08}, we find that the ratios are in the range $N$(N\,\textsc{v}) / $N$(O\,\textsc{vi}) = 0.08 -- 0.38 (with a mean and median ratio of $\approx 0.19$), which is of the same order as the solar abundance ratio \citep[(N/O)$_{\odot} \approx 0.13$;][]{Asplund09}. The observed relative abundance ratio is more difficult to explain in the photoionization scenario, since O\,\textsc{vi} has a much higher ionization potential ($\approx 115$ eV) than N\,\textsc{v} ($\approx 77$ eV). We note that the [N/O] abundance ratios in low-metallicity systems are typically sub-solar \citep{Pettini08}, exacerbating this inconsistency further.

However, in the pure recombination scenario sketched roughly above, the column density of N\,\textsc{v} should be very sensitive to the electron density and should be very time-dependent, with a value that initially rises very quickly in the first several hours after the burst and then drops exponentially at about a day, depending on the density and recombination rate coefficients. We cannot entirely exclude this simple scenario based on the current data. GRB\,120327A is the only GRB in our sample with two time-separated measurements of N\,\textsc{v} and shows no significant variability from 2 hours to 30 hours ($1\,\sigma$ difference is 26\% of the first measurement). In most scenarios, we would expect a change of two orders of magnitude over this time. However, since that GRB has only upper limits to the X-ray column density it cannot be excluded that the points straddle a rise and fall centred around 1 day. In addition, our current data do not show a strong correlation with time taking the bursts as an ensemble. Time-series observations of N\,\textsc{v} in the future would be very valuable to completely rule out the simple recombination-dominated hypothesis, since for most reasonable H\,\textsc{ii} region electron densities, we would expect the column density of N\,\textsc{v} to rise strongly with time, such that a substantial fraction of nitrogen atoms are recombined at $10^5$\,s ($\approx 1$ day) compared to $10^4$\,s, easily ruling out (or confirming) a recombination-dominant scenario. The recombination outlined above excludes a contribution to the high-ionization column densities from photoionization by the GRB afterglow as suggested by \cite{Prochaska08b}. This effect almost certainly needs to be considered to properly model the time-variation of the column density in the high-ionization lines. As it is, neither the recombination or the photoionization scenarios are in agreement with the observed non-varying N\,\textsc{v} abundance and more time-resolved data is required to resolve this issue in both models.

\subsection{Hot gas in GRB host galaxy halos}

There is now positive evidence that the N\,\textsc{v} (and likely also O\,\textsc{vi}) absorption features seen in GRB afterglows originate in the circumburst medium of GRBs. We note, however, that these ions are also observed in the Galactic halo\footnote{Here we use the term Galactic halo to distinguish the hot gas surrounding the host from the neutral ISM of the galaxy.} \citep{Savage97,Sembach03,Wakker12} but rarely with column densities as large as that observed for the GRB sample. \cite{Prochaska08b} use this argument, together with theoretical predictions, to exclude the galactic halo gas as the origin of N\,\textsc{v} and O\,\textsc{vi} absorption in GRB afterglows. The presence of N\,\textsc{v} and O\,\textsc{vi} absorption (though rare) in DLAs and sub-DLAs in the line-of-sight toward quasars \citep{Fox07a,Fox09}, however, likely originate from their galactic halos. Similar low column densities, as in the local Galactic halo, are observed for these samples. Moreover, the column density of N\,\textsc{v} in these quasar absorbers is found to be correlated with the column density and velocity width of C\,\textsc{iv} \citep{Fox09} which supports a circumgalactic origin.

The fact that the velocity width of C\,\textsc{iv} is generally more extended than the neutral gas profiles \citep[e.g.][]{Ledoux98,Wolfe00,Fox07b} is in good agreement with a model where C\,\textsc{iv} originates in the hot gas constituting the galactic halos \citep{Maller03}. That the same is true for C\,\textsc{iv} and Si\,\textsc{iv} absorption observed in GRB afterglows was motivated by e.g. \cite{Fox08} who found that the velocity widths of these ions are similar to those observed for DLAs toward quasars \citep{Fox07b}, with multiple components in a velocity range of several hundred km s$^{-1}$ around $z_{\mathrm{GRB}}$ (although still with a larger average C\,\textsc{iv} column density). 

With the advent of space borne UV spectroscopy, interstellar absorption lines of highly-ionized species (C\,\textsc{iv}, Si\,\textsc{iv}, N\,\textsc{v}, and O\,\textsc{vi}) have also been detected in several Galactic sight lines towards OB-type stars \citep[e.g.,][]{Black80}. These detections demonstrate the presence of highly ionized "coronal" gas in the disk (and halo) of the Milky Way as well as in other nearby galaxies \citep{Lehner11}. The absorption by C\,\textsc{iv} and Si\,\textsc{iv} exhibits broad and narrow components, while only broad components are seen in N\,\textsc{v} and O\,\textsc{vi}. The narrow components are associated with the vicinities of O-type stars and their surrounding supershells. The broad components trace collisionally ionized gas ($T > 10^{5}$~K) in the hot ISM. 

It is not possible to robustly measure the column densities of C\,\textsc{iv} and Si\,\textsc{iv} for the GRBs in our sample due to heavy saturation. We confirm, however, that the velocity widths of these ions are generally much larger than the line profiles of the low-ionization lines and N\,\textsc{v}. Moreover, the peak optical depth of C\,\textsc{iv} and Si\,\textsc{iv} is not aligned with N\,\textsc{v}, except in a few cases where all three ions are aligned with $z_{\mathrm{GRB}}$. In conclusion, while C\,\textsc{iv} and Si\,\textsc{iv} absorption likely originate in the galactic halos of DLAs and sub-DLAs toward quasars and surrounding GRB hosts, N\,\textsc{v} and O\,\textsc{vi} absorption is caused by gas at two distinct regions in the two cases; one from the hot halo gas around DLA and sub-DLAs toward quasars and the other from the gas in the vicinity of the GRB explosion site (though also with a possible small contribution from the hot halo gas).


\section{Concluding remarks} \label{sec:conc}

We have performed a survey for high-ionization absorption in 18 afterglow spectra of long-duration GRBs at $z\approx 1.7 - 6$ observed with VLT/X-shooter. We surveyed the high-ionization species of N\,\textsc{v}, O\,\textsc{vi}, C\,\textsc{iv}, Si\,\textsc{iv}, S\,\textsc{iv}, S\,\textsc{vi} and P\,\textsc{v}, but focus on N\,\textsc{v} since it is the least saturated and blended of these ions. Out of the 18 bursts, we detect N\,\textsc{v} in 12 of the cases, all having large column densities for which we found average and median values of $\log N$(N\,\textsc{v})$_{\mathrm{avg}} = 14.36$ and $\log N$(N\,\textsc{v})$_{\mathrm{med}} = 14.50$, respectively. We place $3\sigma$ upper limits on N\,\textsc{v} for the remaining six bursts and show that the non-detections are likely either related to the intrinsic or observed faintness of the GRBs, resulting in poor signal-to-noise ratios at the spectral region covering the N\,\textsc{v} doublet. 

By comparing the strength of N\,\textsc{v} to the neutral gas properties in individual bursts such as metallicity, H\,\textsc{i} column density and dust depletion, both from our sample and previous samples from the literature, we found no correlation between any of these measurements. This suggests that the gas responsible for the N\,\textsc{v} absorption is unrelated to the properties of the neutral gas in the GRB host galaxy ISM. We found, however, that for GRBs with H\,\textsc{i} column densities below $\approx 10^{21}$ cm$^{-2}$, the line profile of N\,\textsc{v} is offset by $\delta v\approx - 100$\,km\,s$^{-1}$ relative to $z_{\mathrm{GRB}}$, whereas for higher H\,\textsc{i} column densities the N\,\textsc{v} gas is kinematically "cold", where $\lvert\delta v\rvert \lesssim 30$\,km\,s$^{-1}$. This suggests a scenario where the natal H\,\textsc{ii} region of the GRB containing the N\,\textsc{v} gas is confined by the pressure from the surrounding neutral gas in the ISM.

It has been argued that since we observe soft X-ray absorption in excess of Galactic in the majority of GRB afterglows, there must be a significant column density of highly ionized gas within GRB host galaxies. Based on our sample and two previous samples from the literature we found that the X-ray derived metal column density, $N_{\mathrm{H,X}}$, is positively correlated (at $2\sigma$ significance) with the column density of N\,\textsc{v}. Since the line profiles of N\,\textsc{v} are also found to be coincident in velocity with the UV-pumped fine-structure absorption lines, we affirm that both measurements likely trace the column density of metals in the circumburst medium of the GRB. We then demonstrate that a simple model, where the initial prompt emission from the GRB has stripped the electrons from the surrounding gas entirely, can reproduce the observed column densities of N\,\textsc{v} and O\,\textsc{vi} via recombination. We here assumed that the gas is fully ionized out to $\approx 10$ pc within the first few seconds after the GRB, and that the gas has $T=10^4$\,K and an electron density of $n_e = 2\times 10^4$\,cm$^{-3}$ typical for H\,\textsc{ii} regions. 

We would like to encourage future and more detailed modelling of the proposed recombination scenario. Such a model would also have to include the later contribution from photoionization by the afterglow to the observed column densities. A complementary approach would be to measure variations in the column density of N\,\textsc{v} as a function of time in GRB afterglow spectra. We searched for such a temporal variation in the first and second epoch VLT/X-shooter spectra of GRB\,120327A (obtained at $\Delta t_{\mathrm{obs}} \approx 2.1$ and 30.6\,h), but the equivalent width of N\,\textsc{v} is not observed to vary with statistical significance. A similar result was found by \cite{Prochaska08b} for GRBs\,050730 and 050820. Finally, if the recombination scenario proposed here is true we would expect to observe strong emission features from the transitions of e.g N\,\textsc{vii} to N\,\textsc{vi} and N\,\textsc{vi} to N\,\textsc{v}. These line emissions might be possible to observe for low-$z$ GRBs but will require a telescope that is sensititive in the near- and far-ultraviolet.

\section*{Acknowledgements}
We would like to thank the referee for a clear and thorough report, which greatly improved the presentation of the results in the paper. 
We also wish to thank K. Wiersema and P. M\o ller for enlightening discussions.	
KEH and PJ acknowledge support by a Project Grant (162948--051) from The Icelandic Research Fund. JK acknowledges financial support from the Danish Council for Independent Research (EU-FP7 under the Marie-Curie grant agreement no. 600207) with reference DFF--MOBILEX--5051--00115. JJ acknowledges support from NOVA and NWO-FAPESP grant for advanced instrumentation in astronomy. AG acknowledges the financial support from the Slovenian Research Agency (research core funding No. P1-0031 and project grant No. J1-8136). AdUP, CCT and ZC acknowledge support from the Spanish Ministry of Economy and Competitivity under grant number AYA 2014-58381-P. AdUP and CCT acknowledge support from Ramon y Cajal fellowships (RyC-2012-09975 and RyC-2012-09984). AdUP acknowledges support from a grant from the BBVA foundation for researchers and cultural creators. ZC acknowledges support from the Juan de la Cierva Incorporaci\'on fellowship IJCI-2014-21669 and from the Spanish research project AYA 2014-58381-P. RS-R acknowledges support from ASI (Italian Space Agency) through the Contract n. 2015-046-R.0 and from European Union Horizon 2020 Programme
under the AHEAD project (grant agreement n. 654215).



\bibliographystyle{mnras}
\bibliography{ref}


\appendix

\section{Notes on individual GRBs}

\subsection{GRB\,090809 at $z=2.73706$}

\begin{figure} 
	\centering
	\epsfig{file=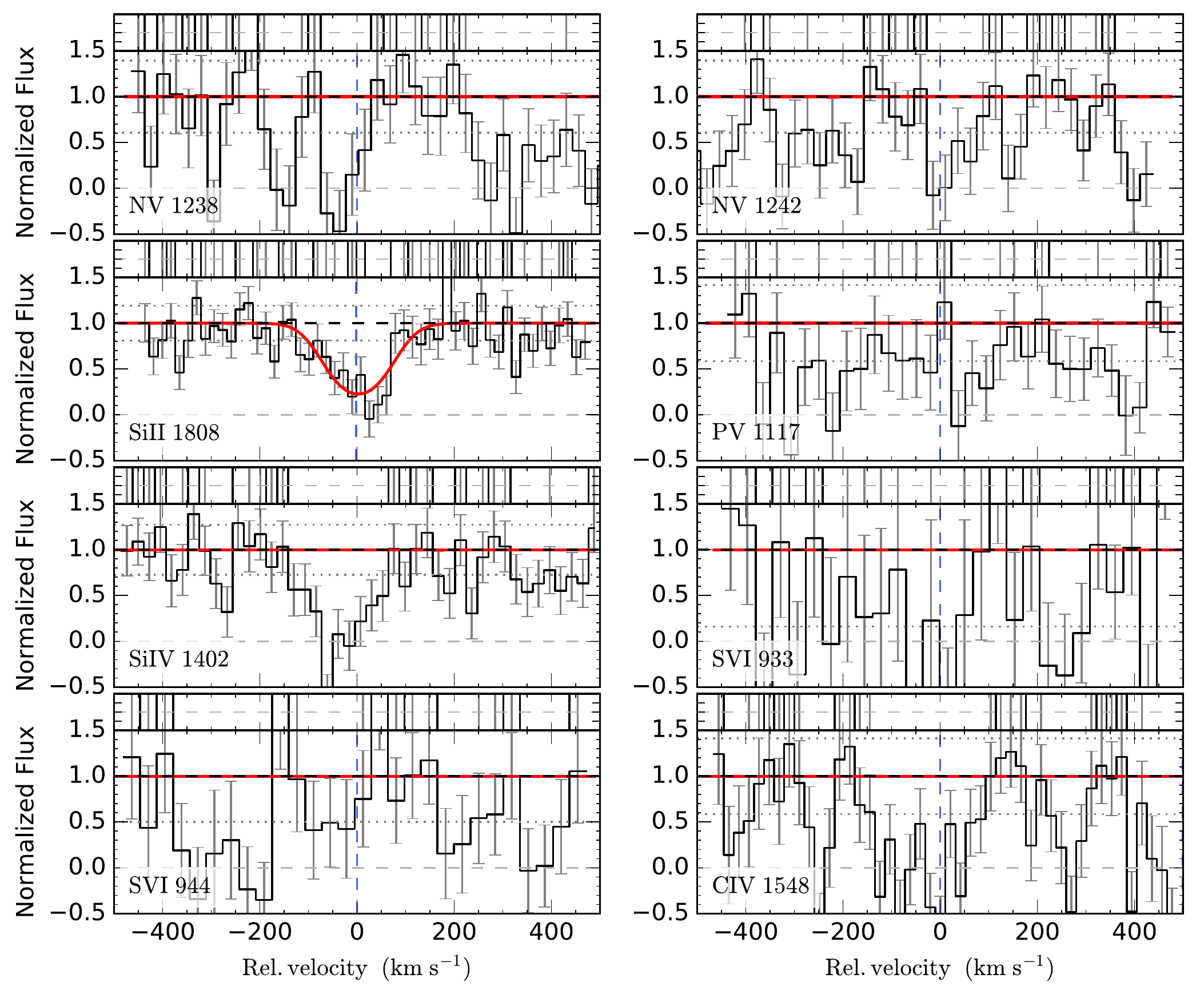,page=1,width=8.9cm}
	\epsfig{file=GRB090809_hion.pdf,page=2,width=8.9cm}
	\caption{GRB090809.}
	\label{fig:090809_hion}
\end{figure}

The spectrum of GRB\,090809A was secured 10.2 hours after trigger with an acquisition magnitude of 21.0 mag. The H\,\textsc{i} column density is adopted from Tanvir et al. (submitted) and was used for the determination of the metallicity. From the fit to Si\,\textsc{ii}\,$\lambda$\,1808 Th\"one et al. (in preparation) find [Si/H] = $-0.76\pm 0.20$. We detect high-ionization absorption features from C\,\textsc{iv} and Si\,\textsc{iv} and measure an upper limit for N\,\textsc{v} of $\log N$(N\,\textsc{v}) < 14.43 (see Fig.~\ref{fig:090809_hion}). The spectral data has low S/N at the location of the other surveyed high-ionization species. The detected high-ionization features are consistent with a single velocity component at $\approx z_{\mathrm{GRB}}$. The X-ray derived metal column density used in Sect.~\ref{ssec:cbm} for this burst is from the \textit{Swift}/XRT database\footnote{\url{http://www.swift.ac.uk/xrt_spectra/00359530/}}.

\subsection{GRB\,090926A at $z=2.10694$}

\begin{figure} 
	\centering
	\epsfig{file=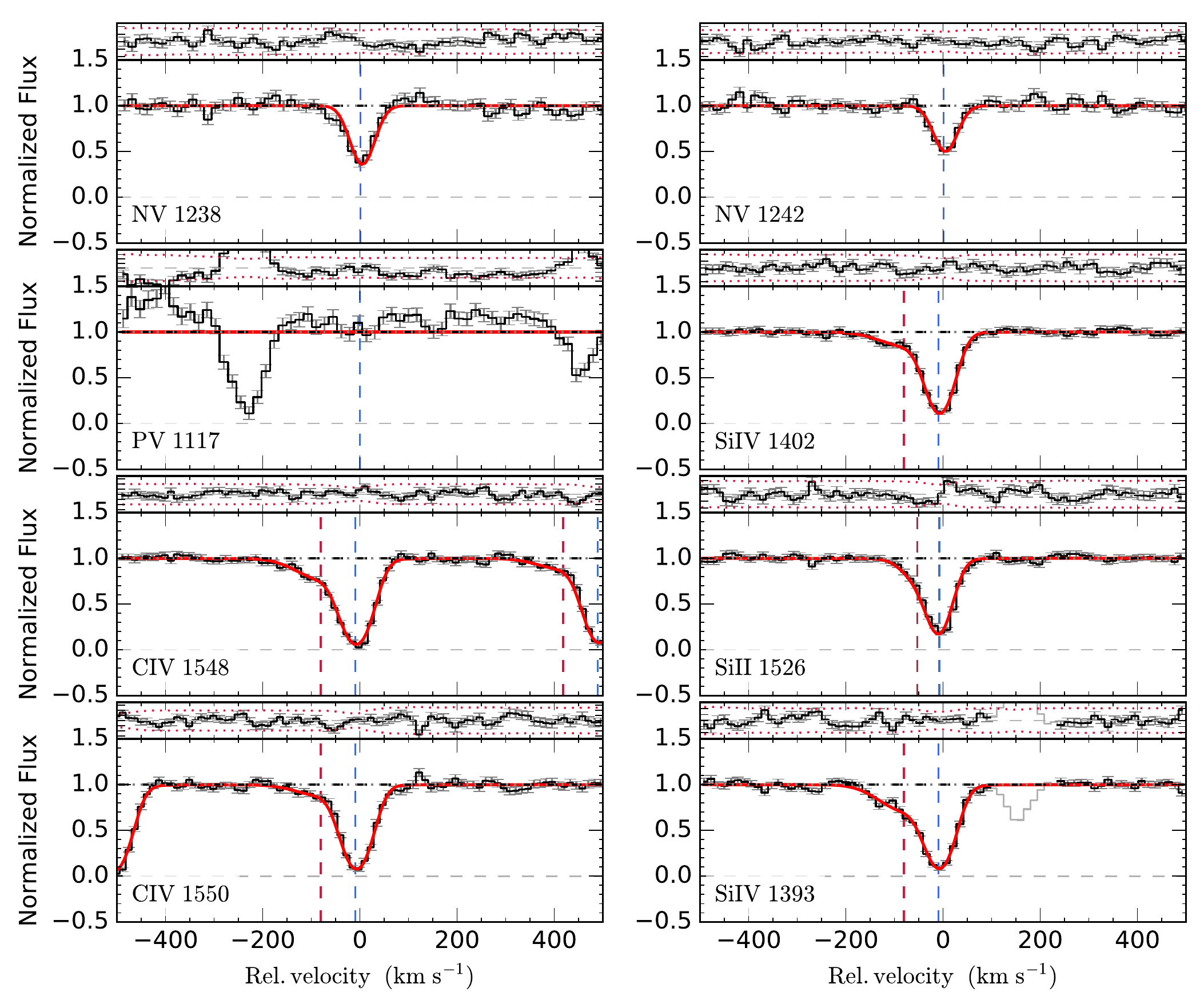,page=1,width=8.9cm}
	\epsfig{file=GRB090926A_hion.pdf,page=2,width=8.9cm}
	\caption{GRB090926A.}
	\label{fig:090926A_hion}
\end{figure}

The spectrum of GRB\,090926A was secured 22.0 hours after trigger with an acquisition magnitude of 17.9 mag. The H\,\textsc{i} column density and metallicity are adopted from \cite{Delia10}. We detect high-ionization absorption features from N\,\textsc{v}, C\,\textsc{iv}, Si\,\textsc{iv}, S\,\textsc{iv} and O\,\textsc{vi} (see Fig.~\ref{fig:090926A_hion}). The absorption lines from S\,\textsc{vi}\,$\lambda\lambda$\,933,944 are located outside the spectral coverage of the UVB arm at this redshift. There is a strong absorption feature at $-200$\,km\,s$^{-1}$ relative to $z_{\mathrm{GRB}}$ at the position of P\,\textsc{v}\,$\lambda$\,1117, but since we do not detect P\,\textsc{v}\,$\lambda$\,1128 at the same redshift this is likely an unrelated line. The strongest component of the high-ionization features is located at $z_{\mathrm{GRB}}$, where C\,\textsc{iv} and Si\,\textsc{iv} show an additional component at $\approx -75$\,km\,s$^{-1}$ relative to $z_{\mathrm{GRB}}$. The velocity components of N\,\textsc{v} are best fit following the low-ionization lines, represented by Si\,\textsc{ii}\,$\lambda$\,1526 in the figure. The line profiles of O\,\textsc{vi} show two strong components, one at $z_{\mathrm{GRB}}$, the other at $\approx -200$\,km\,s$^{-1}$ relative to $z_{\mathrm{GRB}}$. The X-ray derived metal column density used in Sect.~\ref{ssec:cbm} for this burst is from Zafar et al. (submitted).

\subsection{GRB\,100425A at $z=1.75640$}

\begin{figure}
	\centering
	\epsfig{file=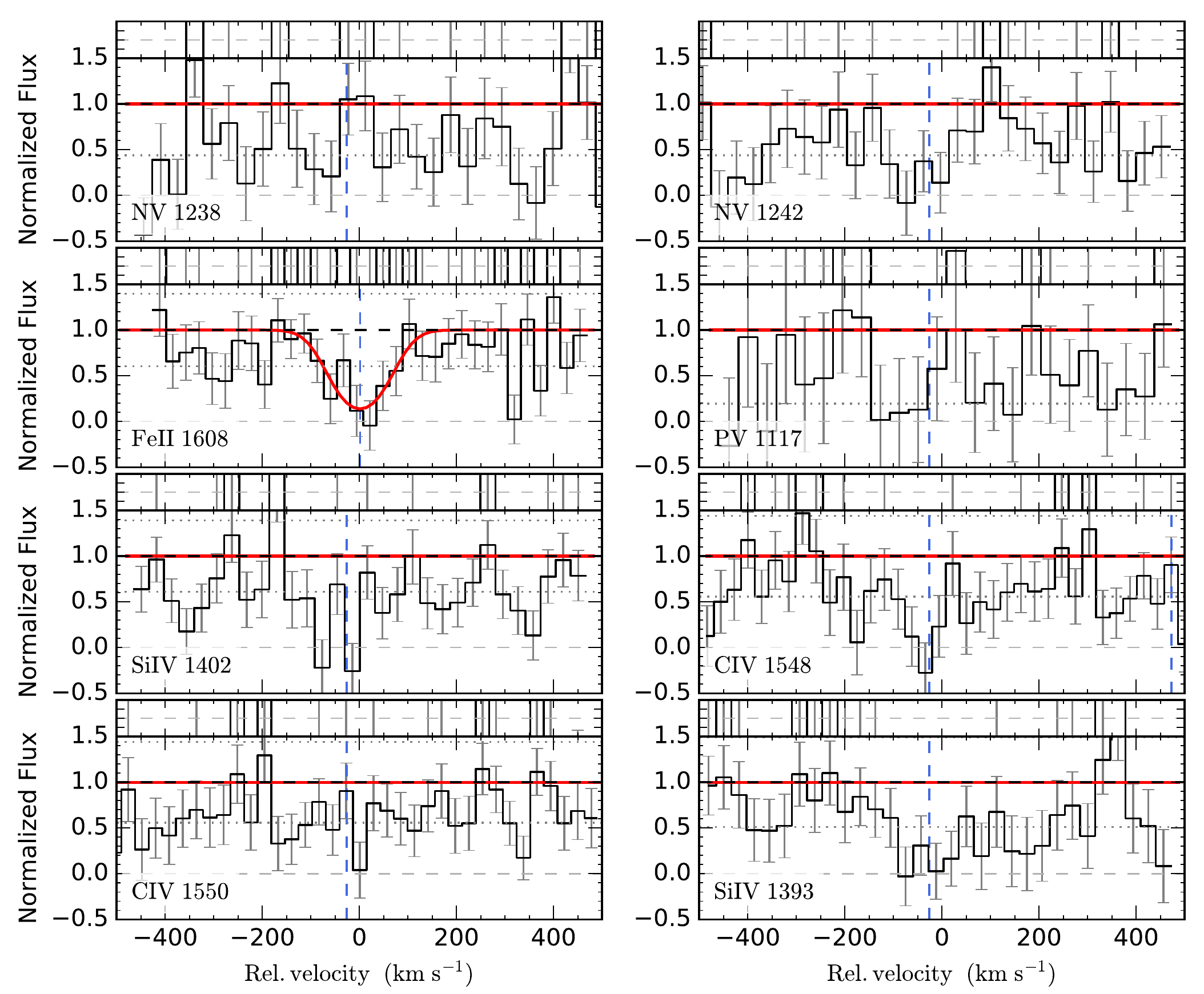,width=8.9cm}
	\caption{GRB100425A. The spectrum has been binned by a factor of 2.}
	\label{fig:100425A_hion}
\end{figure}

The spectrum of GRB\,100425A was secured 4.0 hours after trigger with an acquisition magnitude of 20.6 mag. The H\,\textsc{i} column density and metallicity are adopted from \cite{Cucchiara15}. We detect high-ionization absorption features from C\,\textsc{iv} and Si\,\textsc{iv} (see Fig.~\ref{fig:100425A_hion}) and measure an upper limit for N\,\textsc{v} of $\log N$(N\,\textsc{v}) < 14.63. We caution, however, that all features are located in regions with poor spectral S/N and O\,\textsc{vi}, S\,\textsc{iv}, and S\,\textsc{vi} are outside the spectral coverage of the UVB arm. Because of this we do not report column densities for these transitions. No information on the velocity components are available from C\,\textsc{iv} and Si\,\textsc{iv}, except that the absorption line profiles cover $\approx -100 < z_{\mathrm{GRB}} < +100$\,km\,s$^{-1}$. The X-ray derived metal column density used in Sect.~\ref{ssec:cbm} for this burst is from the \textit{Swift}/XRT database\footnote{\url{http://www.swift.ac.uk/xrt_spectra/00420398/}}.

\subsection{GRB\,111008A at $z=4.99146$}

\begin{figure}
	\centering
	\epsfig{file=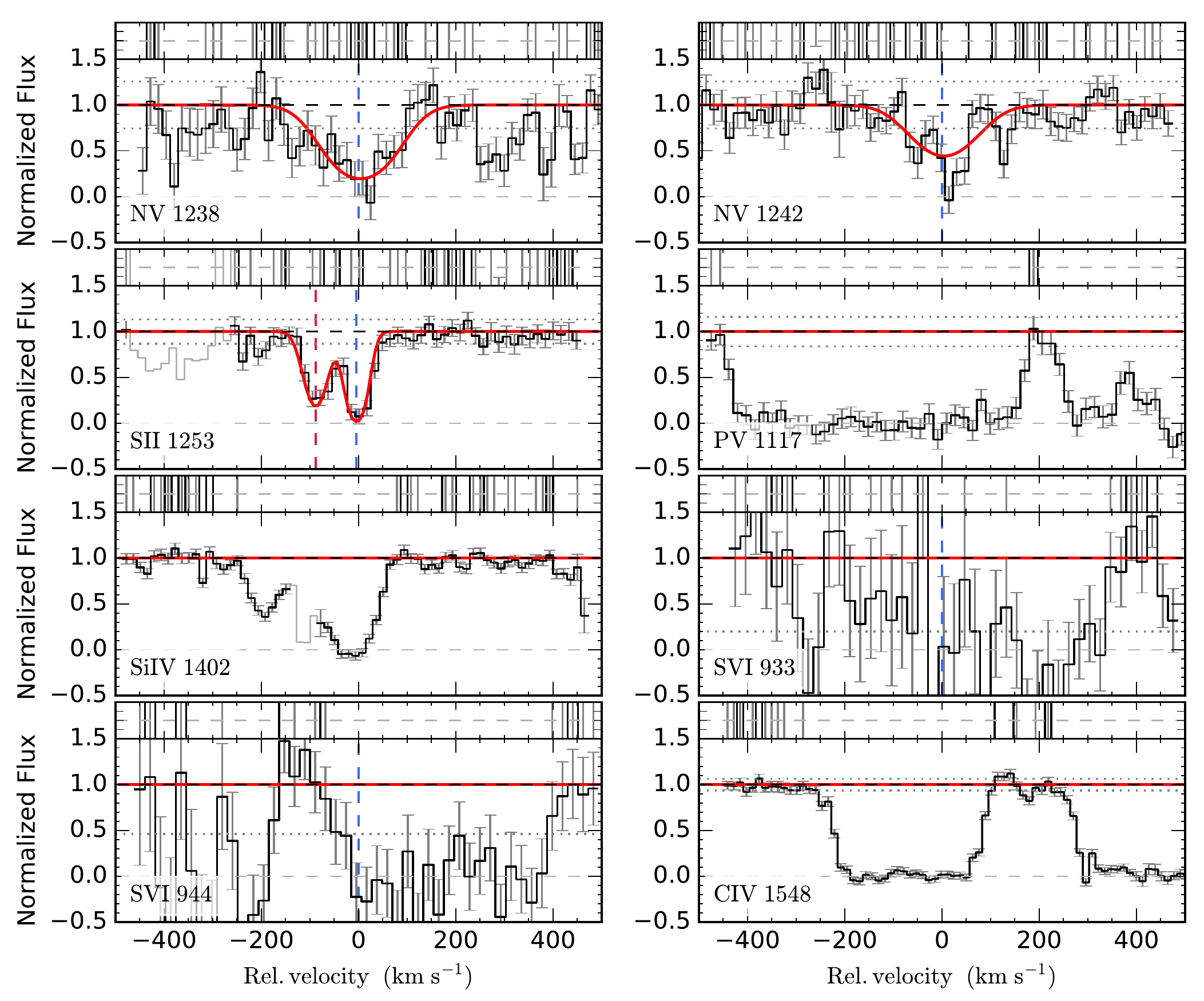,page=1,width=8.9cm}
	\epsfig{file=GRB111008A_hion.pdf,page=2,width=8.9cm}
	\caption{GRB111008A. The spectrum has been binned by a factor of 2.}
	\label{fig:111008A_hion}
\end{figure}

The spectrum of GRB\,111008A was secured 8.5 hours after trigger with an acquisition magnitude of 21.0 mag. The H\,\textsc{i} column density and metallicity are adopted from \cite{Sparre14}. We detect high-ionization absorption features from N\,\textsc{v}, C\,\textsc{iv}, and Si\,\textsc{iv} (see Fig.~\ref{fig:111008A_hion}). The absorption lines from C\,\textsc{iv} and Si\,\textsc{iv} are saturated and partly blended so we do not report column densities for these transitions. We report a non-detection of the remaining surveyed high-ionization species, but caution that they are all located in spectral regions with poor S/N. No information on the velocity components are available from C\,\textsc{iv} and Si\,\textsc{iv}, except that the absorption line profiles cover $\approx -200 < z_{\mathrm{GRB}} < +100$\,km\,s$^{-1}$. N\,\textsc{v} is consistent with a single velocity component at $\approx z_{\mathrm{GRB}}$, but we are only able to infer a lower limit N\,\textsc{v} due to line saturation. The X-ray derived metal column density used in Sect.~\ref{ssec:cbm} for this burst is from Zafar et al. (submitted).

\subsection{GRB\,120119A at $z=1.72883$}

\begin{figure} 
	\centering
	\epsfig{file=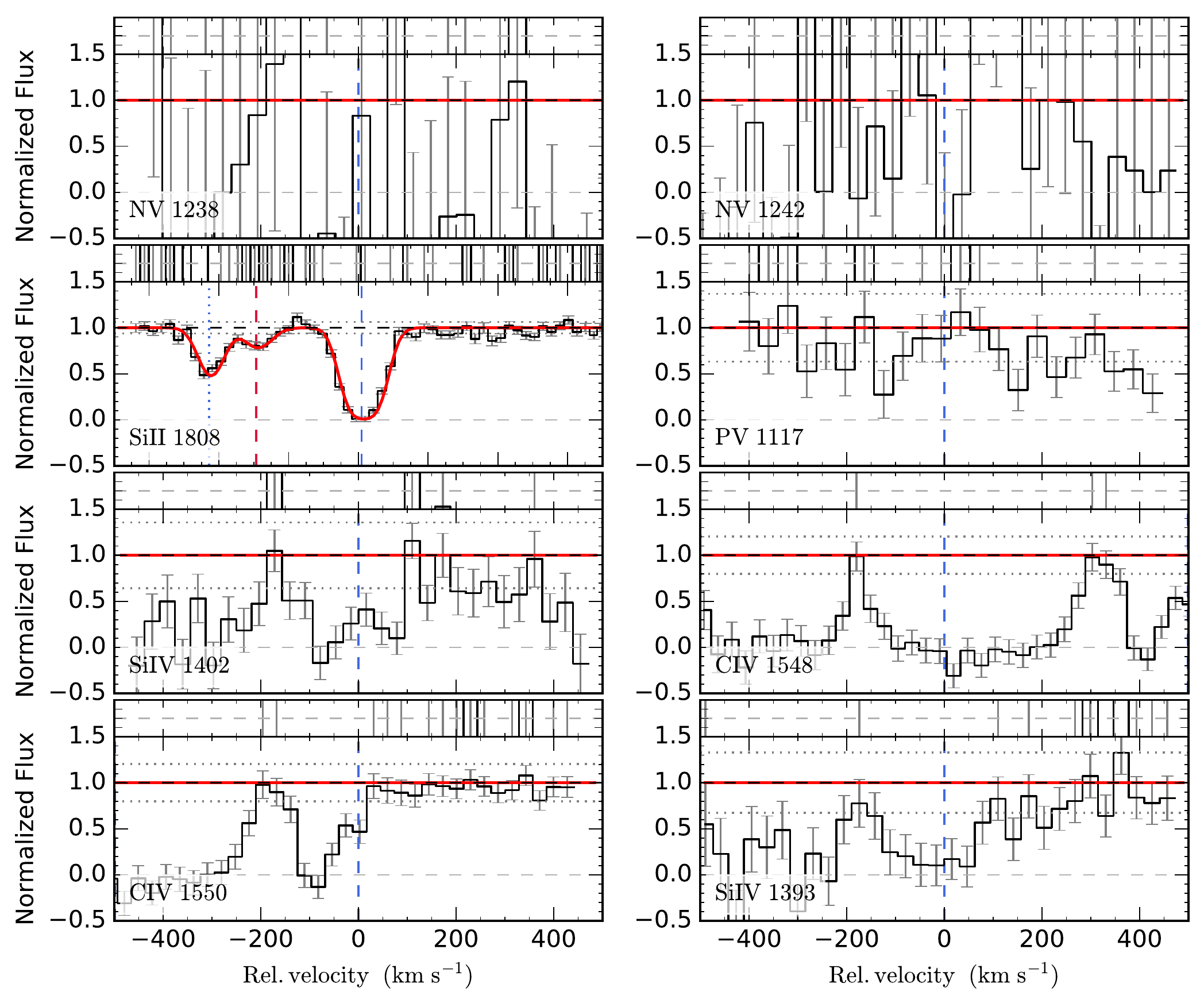,page=1,width=8.9cm}
	\caption{GRB120119A. The spectrum has been binned by a factor of 2.}
	\label{fig:120119A_hion}
\end{figure}

The spectrum of GRB\,120119A was secured 1.4 hours after trigger with an acquisition magnitude of 17.0 mag. The H\,\textsc{i} column density is adopted from Tanvir et al. (submitted) and was used for the determination of the metallicity. Th\"one et al. (in preparation) measure [Si/H] = $-1.43\pm 0.21$ from a fit to S\,\textsc{ii}\,$\lambda\lambda$\,1808. We detect high-ionization absorption features from C\,\textsc{iv} and Si\,\textsc{iv} (see Fig.~\ref{fig:120119A_hion}) and measure an upper limit for N\,\textsc{v} of $\log N$(N\,\textsc{v}) < 15.15. We caution, however, that N\,\textsc{v} is located in a region with poor spectral S/N and O\,\textsc{vi}, S\,\textsc{iv}, and S\,\textsc{vi} are outside the spectral coverage of the UVB arm. The absorption lines from C\,\textsc{iv} and Si\,\textsc{iv} are heavily saturated and partly blended so we do not report column densities for these transitions. We report a non-detection of the remaining surveyed high-ionization species, but caution that they are all located in spectral regions with poor S/N. No information on the velocity components are available from C\,\textsc{iv} and Si\,\textsc{iv}, except that the absorption line profiles cover $\approx -700 < z_{\mathrm{GRB}} < 0$\,km\,s$^{-1}$. The X-ray derived metal column density used in Sect.~\ref{ssec:cbm} for this burst is from Zafar et al. (submitted).

\subsection{GRB\,120327A at $z=2.81482$}

The spectrum of GRB\,120317A was secured 2.1 hours after trigger with an acquisition magnitude of 18.8 mag. The H\,\textsc{i} column density, metallicity and high-ionization absorption lines are adopted from \cite{Delia14}. They detect all the surveyed high-ionization absorption features except for S\,\textsc{iv} and are consistent with being located at the central velocity component (out of three extending from $\approx \pm100$\,km\,s$^{-1}$ relative to $z_{\mathrm{GRB}}$) of the low-ionization species. They measure $\log N$(N\,\textsc{v}/cm$^{-2}$) = $13.56\pm 0.03$ which was used in our analysis. The X-ray derived metal column density used in Sect.~\ref{ssec:cbm} for this burst is from the \textit{Swift}/XRT database\footnote{\url{http://www.swift.ac.uk/xrt_spectra/00518731/}}.

\subsection{GRB\,120815A at $z=2.35820$}

\begin{figure} 
	\centering
	\epsfig{file=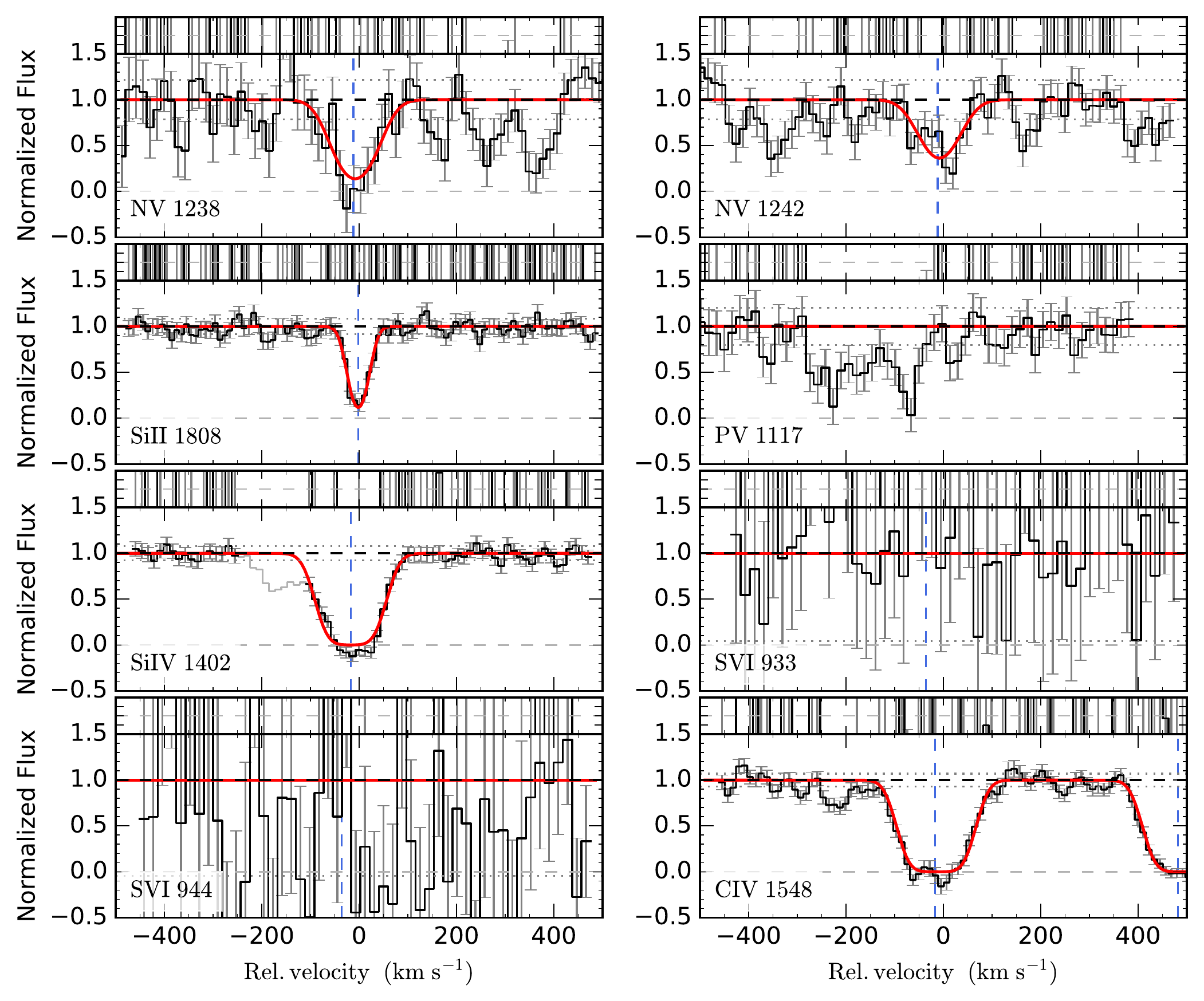,page=1,width=8.9cm}
	\epsfig{file=GRB120815A_hion.pdf,page=2,width=8.9cm}
	\caption{GRB120815A.}
	\label{fig:120815A_hion}
\end{figure}

The spectrum of GRB\,120815A was secured 1.69 hours after trigger with an acquisition magnitude of 18.9 mag. The H\,\textsc{i} column density, metallicity and X-ray derived metal column density used in Sect.~\ref{ssec:cbm} for this burst are adopted from \cite{Kruehler13}. We detect high-ionization absorption features from N\,\textsc{v}, C\,\textsc{iv}, and Si\,\textsc{iv} (see Fig.~\ref{fig:120815A_hion}). The absorption lines from C\,\textsc{iv} and Si\,\textsc{iv} are saturated and the derived column densities should only be considered as lower limits. We report a non-detection of the remaining surveyed high-ionization species, but caution that they are all located in spectral regions with poor S/N. The high-ionization line profiles are best fit with two velocity components at $z_{\mathrm{GRB}}$ and $\approx -23$\,km\,s$^{-1}$ relative to $z_{\mathrm{GRB}}$, similar to the low-ionization absorption lines, represented by Si\,\textsc{ii}\,$\lambda$\,1808 in the figure.

\subsection{GRB\,120909A at $z=3.92882$}

\begin{figure} 
	\centering
	\epsfig{file=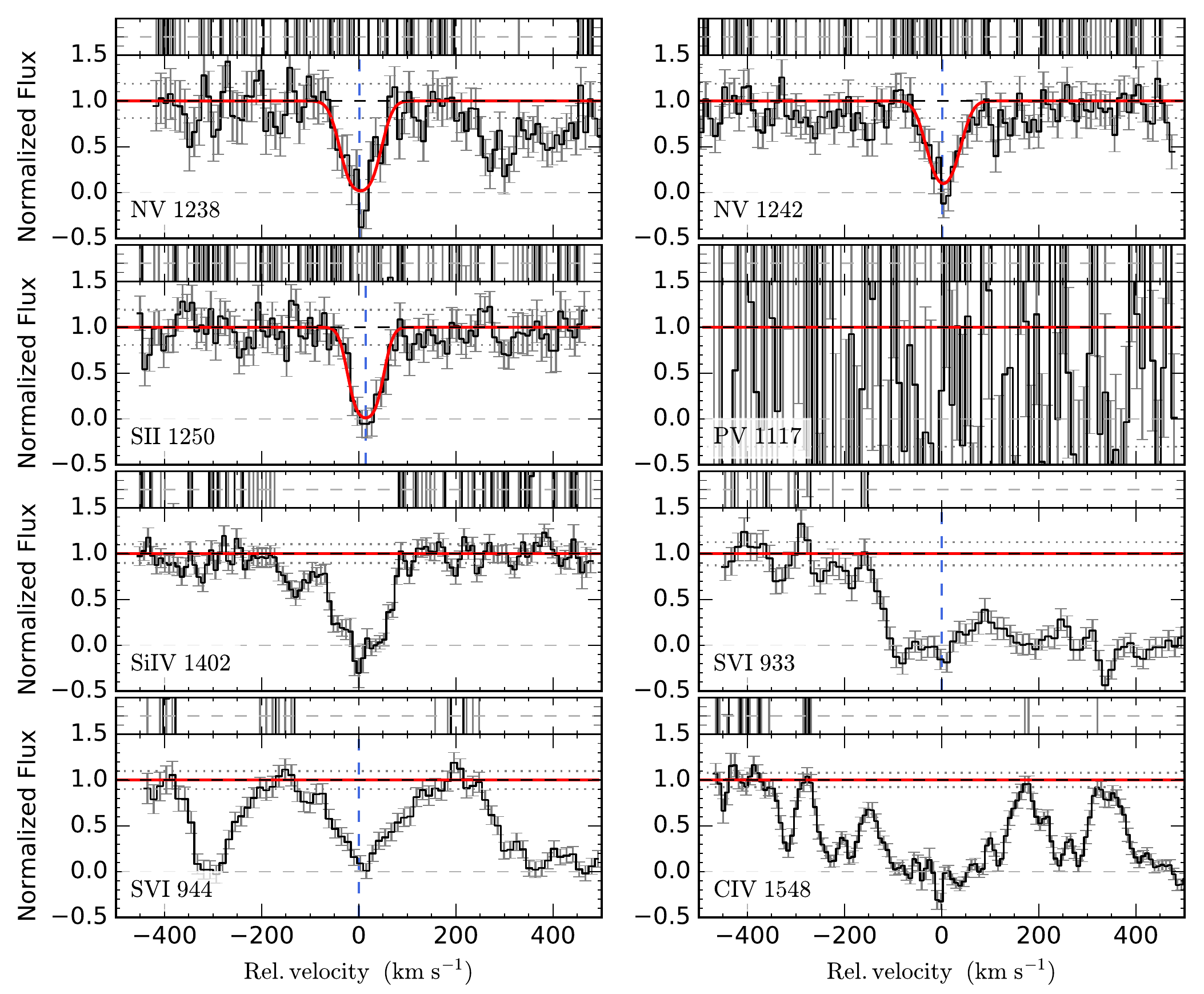,page=1,width=8.9cm}
	\epsfig{file=GRB120909A_hion.pdf,page=2,width=8.9cm}
	\caption{GRB120909A.}
	\label{fig:120909A_hion}
\end{figure}

The spectrum of GRB\,120909A was secured 1.7 hours after trigger using the rapid-response mode with an acquisition magnitude of 21.0 mag. The H\,\textsc{i} column density is adopted from Tanvir et al. (submitted) and was used for the determination of the metallicity. Th\"one et al. (in preparation) measure [S/H] = $-0.66\pm 0.11$ from a combined fit to S\,\textsc{ii}\,$\lambda\lambda$\,1250,1253. We detect high-ionization absorption features from N\,\textsc{v}, C\,\textsc{iv}, Si\,\textsc{iv}, S\,\textsc{iv}, and S\,\textsc{vi} (see Fig.~\ref{fig:120909A_hion}). We only fit Voigt profiles to N\,\textsc{v} and Si\,\textsc{iv} since the other lines are either blended or saturated.
P\,\textsc{v} is located in spectral region with poor S/N and O\,\textsc{vi} is heavily blended with the Ly$\alpha$ forest. The main velocity component of the high-ionization absorption lines is located at $z_{\mathrm{GRB}}$, represented by S\,\textsc{ii}\,$\lambda$\,1250 in the figure. In addition, there is also a small component at $\approx -125$\,km\,s$^{-1}$ relative to $z_{\mathrm{GRB}}$. The X-ray derived metal column density used in Sect.~\ref{ssec:cbm} for this burst is from the \textit{Swift}/XRT database\footnote{\url{http://www.swift.ac.uk/xrt_spectra/00533060/}}.

\subsection{GRB\,121024A at $z=2.30244$}

\begin{figure} 
	\centering
	\epsfig{file=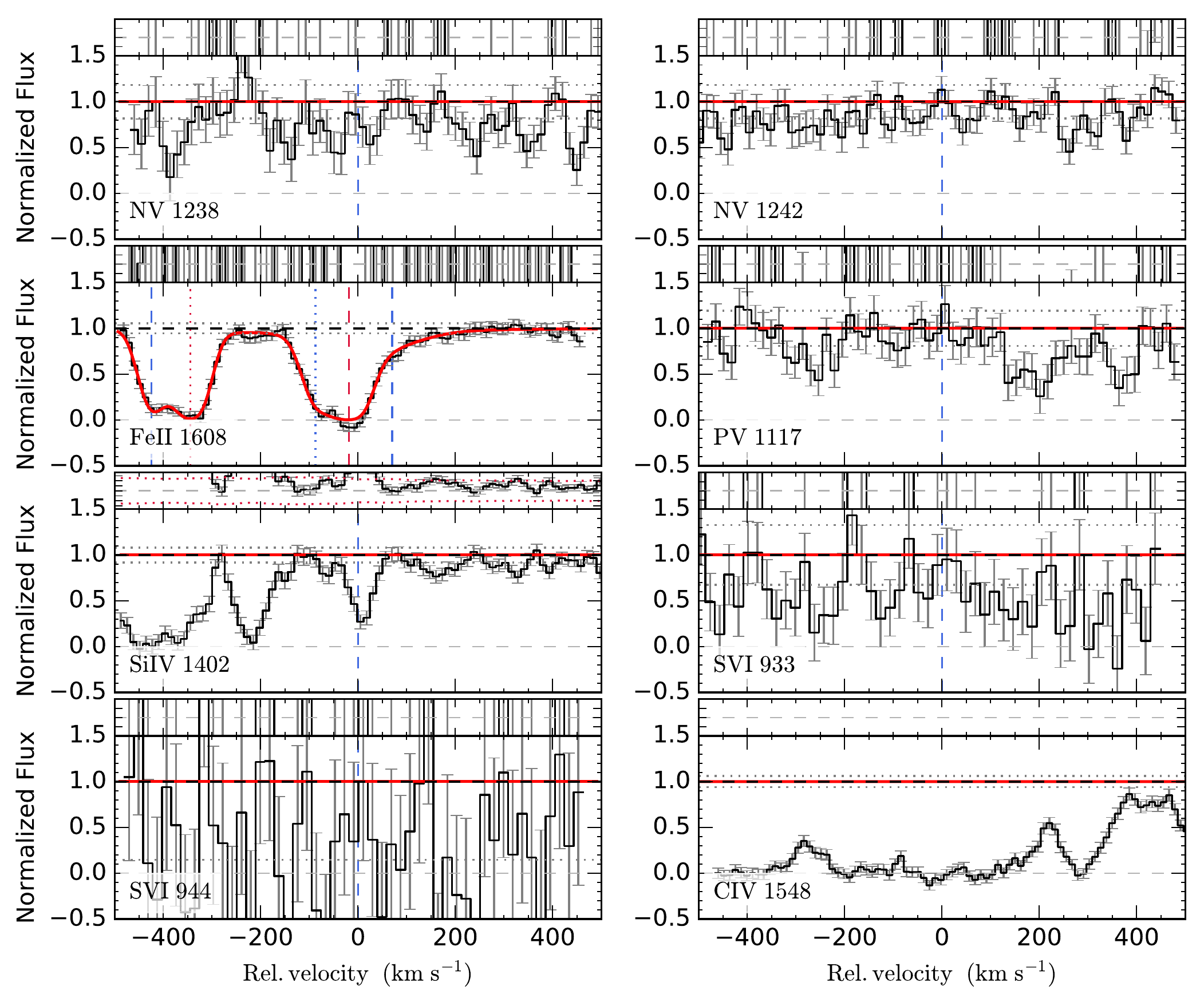,page=1,width=8.9cm}
	\epsfig{file=GRB121024A_hion.pdf,page=2,width=8.9cm}
	\caption{GRB121024A.}
	\label{fig:121024A_hion}
\end{figure}

The spectrum of GRB\,121024A was secured 1.8 hours after trigger with an acquisition magnitude of 20.0 mag. The H\,\textsc{i} column density and metallicity are adopted from \cite{Friis15}. We detect high-ionization absorption from C\,\textsc{iv} and Si\,\textsc{iv} (see Fig.~\ref{fig:121024A_hion}) and measure an upper limit for N\,\textsc{v} of $\log N$(N\,\textsc{v}) < 14.35. Both of these two absorption lines are saturated and the derived column densities should only be considered as lower limits. We report a non-detection of the remaining surveyed high-ionization species, but caution that they are all located in spectral regions with poor S/N. The velocity components of C\,\textsc{iv} and Si\,\textsc{iv} show a weak, ionized cloud at $z_{\mathrm{GRB}}$, and a highly complex distribution at $-600$ to $-100$\,km\,s$^{-1}$ relative to $z_{\mathrm{GRB}}$. For comparison, the low-ionization absorption line profiles, represented by Fe\,\textsc{ii}\,$\lambda$\,1608 in the figure, show two strong components, one at $z_{\mathrm{GRB}}$ and one at $-400$\,km\,s$^{-1}$ relative to $z_{\mathrm{GRB}}$. The X-ray derived metal column density used in Sect.~\ref{ssec:cbm} for this burst is from the \textit{Swift}/XRT database\footnote{\url{http://www.swift.ac.uk/xrt_spectra/00536580/}}.

\subsection{GRB\,130408A at $z=3.75792$}

\begin{figure} 
	\centering
	\epsfig{file=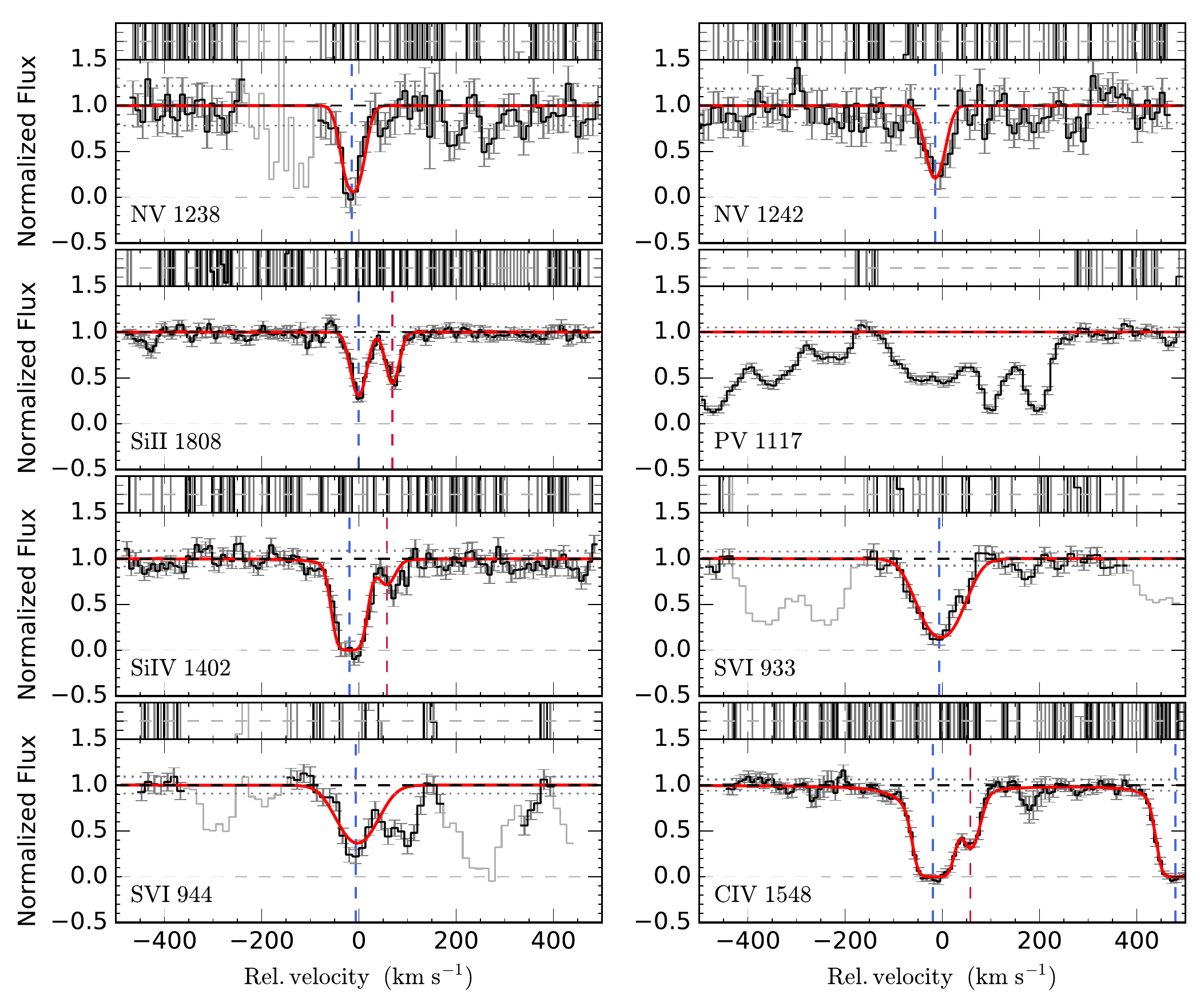,page=1,width=8.9cm}
	\epsfig{file=GRB130408A_hion.pdf,page=2,width=8.9cm}
	\caption{GRB130408A.}
	\label{fig:130408A_hion}
\end{figure}

The spectrum of GRB\,160408A was secured 1.9 hours after trigger with an acquisition magnitude of 20.0 mag. The H\,\textsc{i} column density is adopted from Tanvir et al. (submitted) and was used for the determination of the metallicity. Th\"one et al. (in preparation) measure [Si/H] = $-1.38\pm 0.12$ from a combined fit to Si\,\textsc{ii}\,$\lambda\lambda$\,1526,1808. We detect high-ionization absorption features from N\,\textsc{v}, C\,\textsc{iv}, Si\,\textsc{iv}, and S\,\textsc{vi}. The three features O\,\textsc{vi}, P\,\textsc{v}, and S\,\textsc{iv} are all blended (see Fig.~\ref{fig:130408A_hion}) and therefore we do not attempt to measure their column densities. The absorption line profile of N\,\textsc{v} is consistent with a single velocity component. The velocity components of C\,\textsc{iv}, Si\,\textsc{iv}, and S\,\textsc{vi} follow the same distribution as for the low-ionization line profiles, represented in the figure by Si\,\textsc{ii}\,$\lambda$\,1808. The X-ray derived metal column density used in Sect.~\ref{ssec:cbm} for this burst is from the \textit{Swift}/XRT database\footnote{\url{http://www.swift.ac.uk/xrt_spectra/00553132/}}, but corrected for the measured redshift.

\subsection{GRB\, at $z=5.91278$}

\begin{figure}
	\centering
	\epsfig{file=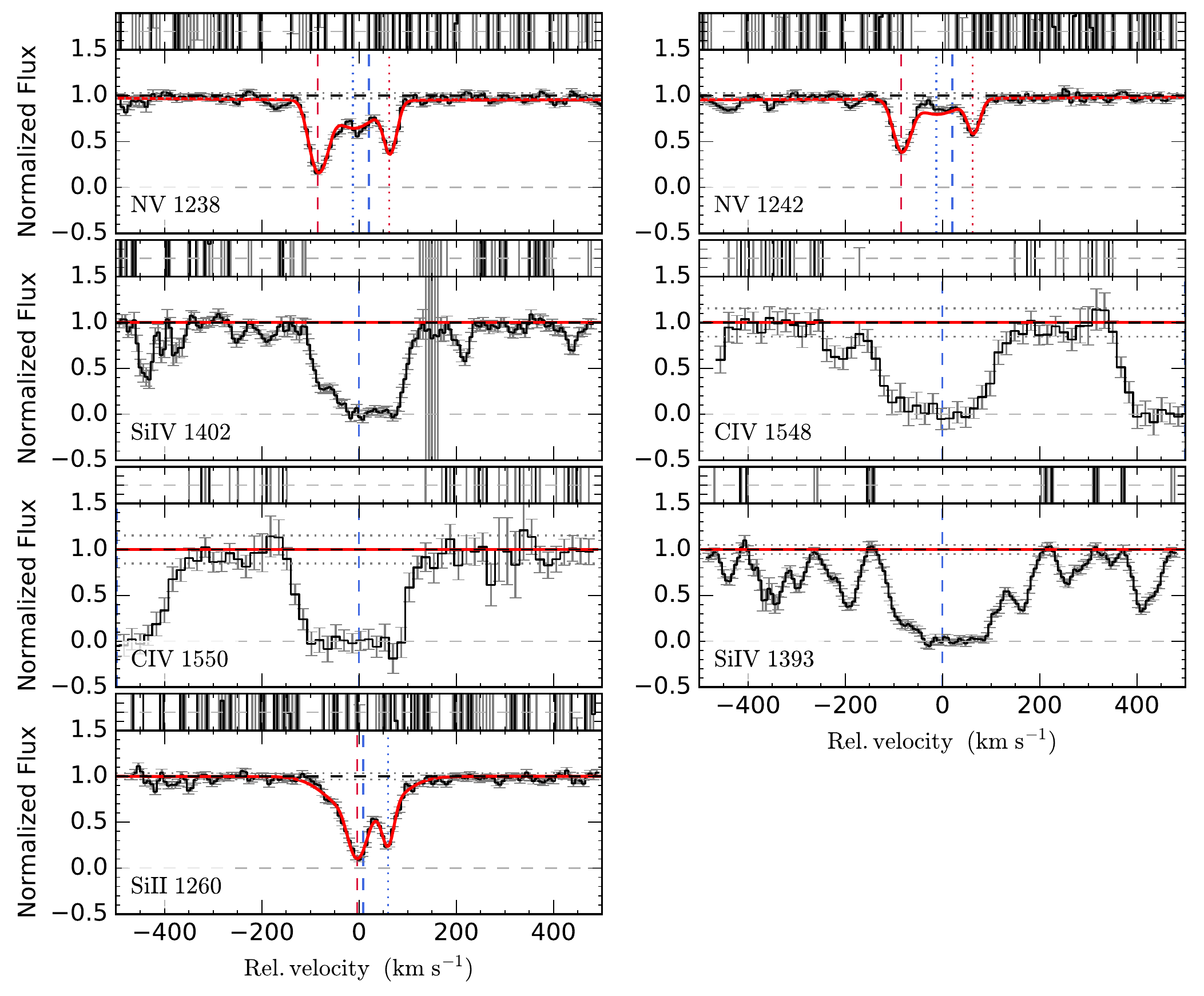,page=1,width=8.9cm}
	\caption{GRB130606A.}
	\label{fig:130606A_hion}
\end{figure}

The spectrum of GRB\,130606A was secured 7.1 hours after trigger with an acquisition magnitude of 19.0 mag. The H\,\textsc{i} column density, metallicity and X-ray derived metal column density used in Sect.~\ref{ssec:cbm} for this burst are adopted from \cite{Hartoog15}. Since the intervening intergalactic medium have a high neutral fraction, the continuum on the blue side of the Ly$\alpha$ absorption feature is completely suppressed. Therefore, we only searched for high-ionization absorption lines redwards of Ly$\alpha$. We detect N\,\textsc{v}, C\,\textsc{iv}, and Si\,\textsc{iv}, all with multiple velocity components (see Fig.~\ref{fig:130606A_hion}). C\,\textsc{iv} and Si\,\textsc{iv} are both saturated so the column densities derived for these two species should only be considered as lower limits. While the velocity components of C\,\textsc{iv} and Si\,\textsc{iv} are consistent with the low-ionization absorption line profiles, represented by Si\textsc{ii}\,$\lambda$\,1260 in the figure, N\,\textsc{v} has two main components at $\approx \pm 100$\,km\,s$^{-1}$ compared to $z_{\mathrm{GRB}}$.

\subsection{GRB\,141028A at $z=2.33327$}

\begin{figure} 
	\centering
	\epsfig{file=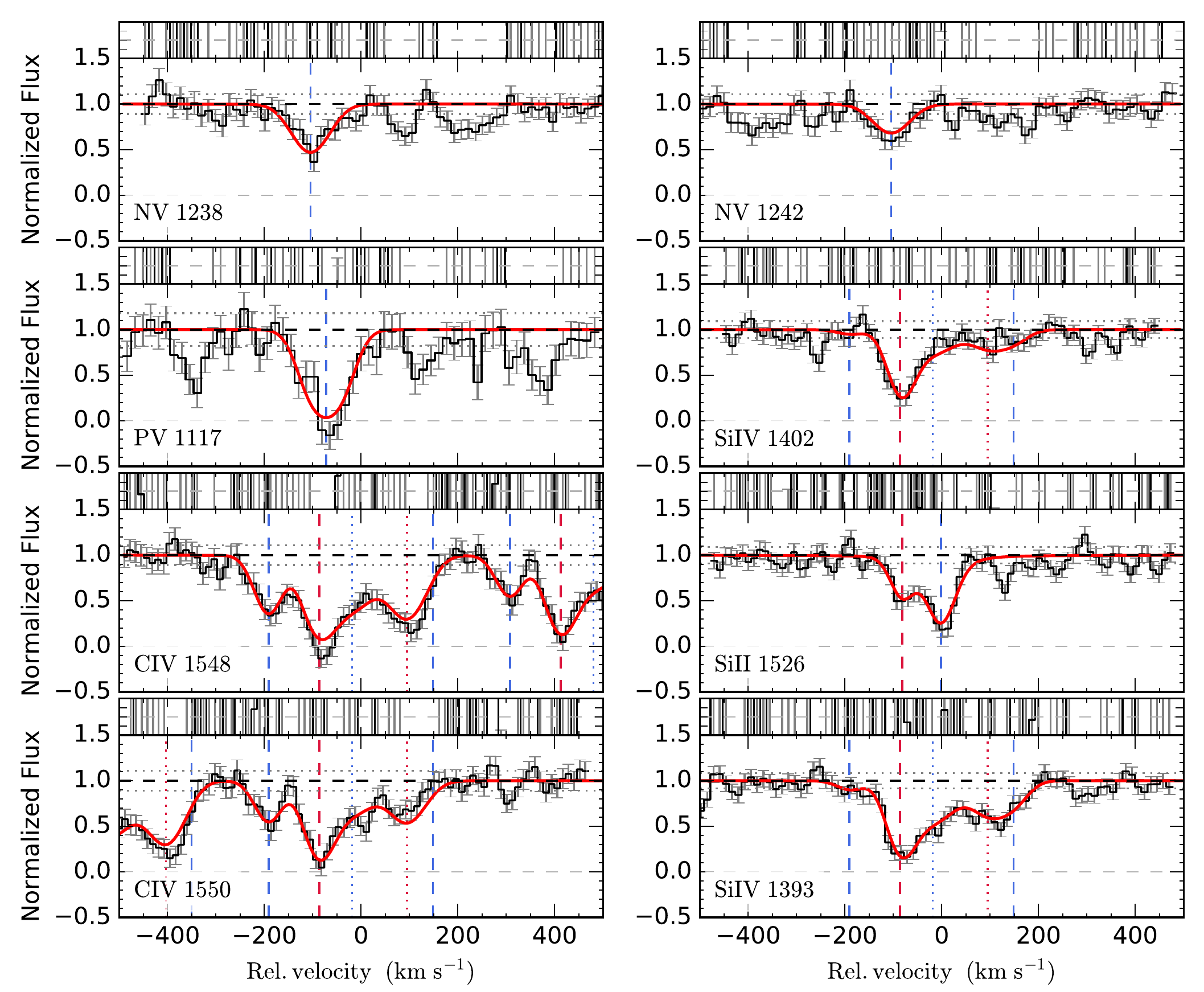,page=1,width=8.9cm}
	\epsfig{file=GRB141028A_hion.pdf,page=2,width=8.9cm}
	\caption{GRB141028A.}
	\label{fig:141028A_hion}
\end{figure}

The spectrum of GRB\,141028A was secured 15.4 hours after trigger with an acquisition magnitude of 20.0 mag. The H\,\textsc{i} column density is adopted from Tanvir et al. (submitted) and was used for the determination of the metallicity. Th\"one et al. (in preparation) measure [Si/H] = $-0.75\pm 0.15$ from a combined fit to  Si\,\textsc{ii}\,$\lambda\lambda$\,1260,1526. We detect high-ionization absorption features from N\,\textsc{v}, C\,\textsc{iv}, Si\,\textsc{iv}, and P\,\textsc{v} (see Fig.~\ref{fig:141028A_hion}). S\,\textsc{vi},$\lambda\lambda$\,933,944 are located bluewards of the spectral coverage of the UVB arm and the region around O\,\textsc{vi} and S\,\textsc{iv} has poor S/N. The main velocity components of the high-ionization absorption line profiles are located at $\approx -100$\,km\,s$^{-1}$ from $z_{\mathrm{GRB}}$. N\,\textsc{v} and P\,\textsc{v} are both well fit with a single velocity component whereas Si\,\textsc{iv} and C\,\textsc{iv} show three and four velocity components, respectively. The X-ray derived metal column density used in Sect.~\ref{ssec:cbm} for this burst is from the \textit{Swift}/XRT database, but corrected for the measured redshift\footnote{\url{http://www.swift.ac.uk/xrt_spectra/00020420/}}.

\subsection{GRB\,141109A at $z=2.99438$}

\begin{figure} 
	\centering
	\epsfig{file=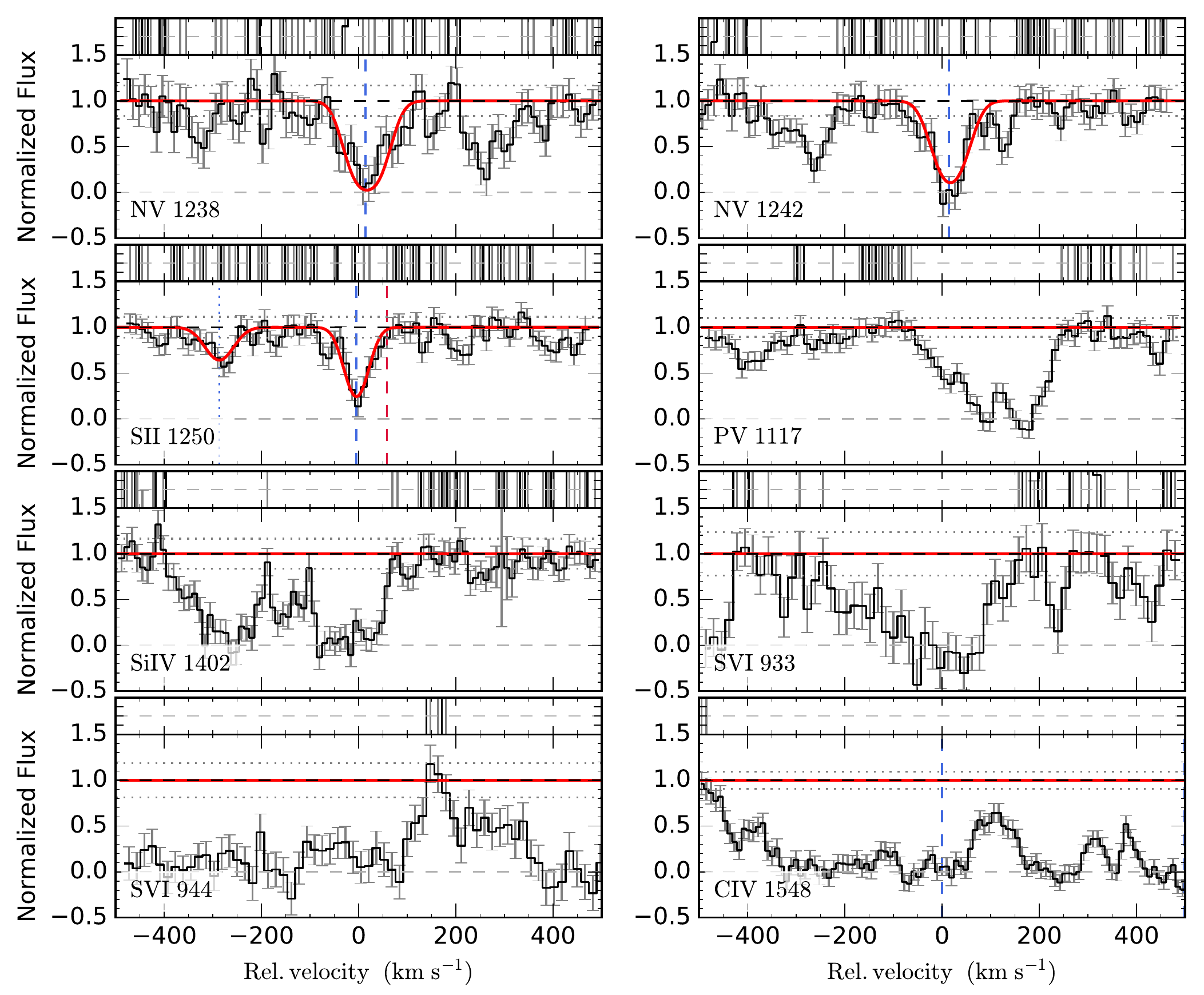,page=1,width=8.9cm}
	\epsfig{file=GRB141109A_hion.pdf,page=2,width=8.9cm}
	\caption{GRB141109A.}
	\label{fig:141109A_hion}
\end{figure}

The spectrum of GRB\,141109A was secured 1.9 hours after trigger with an acquisition magnitude of 19.2 mag. The H\,\textsc{i} column density is adopted from Tanvir et al. (submitted) and was used for the determination of the metallicity. Th\"one et al. (in preparation) measure [Zn/H] = $-1.45\pm 0.21$ from a combined fit to  Zn\,\textsc{ii}\,$\lambda\lambda$\,2026,2062. We detect high-ionization absorption features from N\,\textsc{v}, C\,\textsc{iv}, Si\,\textsc{iv}, S\,\textsc{vi}, P\,\textsc{v}, and S\,\textsc{iv} (see Fig.~\ref{fig:141109A_hion}). Strong absorption from Si\,\textsc{iv}, C\,\textsc{iv}, and P\,\textsc{v} are observed and all features are mildly saturated. The velocity components and column densities of Si\,\textsc{iv} and C\,\textsc{iv} are hard to disentangle since they are both blended and are located in regions with poor S/N. The main component of N\,\textsc{v} is located at $\approx 13$\,km\,s$^{-1}$ from $z_{\mathrm{GRB}}$ and while being best fit by two components we only trust the one central fit. We are also only able to derive a lower limit to the column density for N\,\textsc{v} due to line saturation. The detections of S\,\textsc{vi},$\lambda\lambda$\,933,944, P\,\textsc{v},$\lambda\lambda$\,1117,1128, and S\,\textsc{iv}\,$\lambda$\,1062 are tentative since they all appear blended with the Ly$\alpha$ forest. If the detections of P\,\textsc{v},$\lambda\lambda$\,1117,1128 are real, those are the only features with blueshifted velocity components compared to $z_{\mathrm{GRB}}$, the other are at -400 to 0\,km\,s$^{-1}$. The X-ray derived metal column density used in Sect.~\ref{ssec:cbm} for this burst is from the \textit{Swift}/XRT database\footnote{\url{http://www.swift.ac.uk/xrt_spectra/00618024/}}.

\subsection{GRB\,150403A at $z=2.05707$}

\begin{figure} 
	\centering
	\epsfig{file=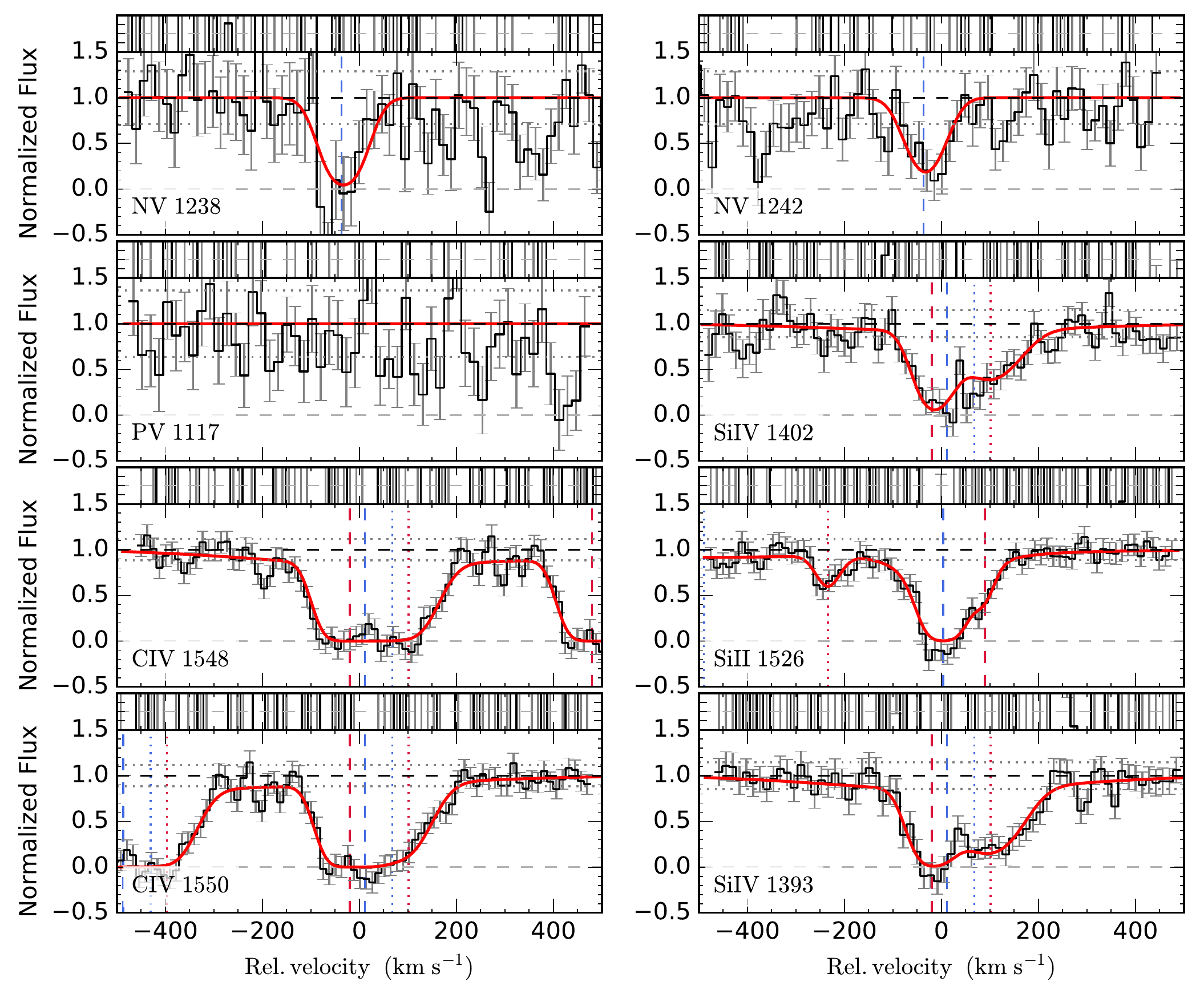,page=1,width=8.9cm}
	\epsfig{file=GRB150403A_hion.pdf,page=2,width=8.9cm}
	\caption{GRB150403A.}
	\label{fig:150403A_hion}
\end{figure}

The spectrum of GRB\,150403A was secured 10.8 hours after trigger with an acquisition magnitude of 19.1 mag. The H\,\textsc{i} column density is adopted from Tanvir et al. (submitted) and was used for the determination of the metallicity. Th\"one et al. (in preparation) measure [S/H] = $-0.80\pm 0.35$ from a combined fit to S\,\textsc{ii}\,$\lambda\lambda$\,1250,1253. We detect high-ionization absorption features from N\,\textsc{v}, C\,\textsc{iv}, and Si\,\textsc{iv} (see Fig.~\ref{fig:150403A_hion}). Strong absorption from Si\,\textsc{iv} and C\,\textsc{iv} are observed and both features are mildly saturated. The four velocity components of both of these line profiles are similar to the low-ionization absorption features, represented in the figure by Si\,\textsc{ii}\,$\lambda$\,1526, and the total line profiles extend from $-200$\,km\,s$^{-1}$ to $+200$\,km\,s$^{-1}$. We find N\,\textsc{v}\,$\lambda\lambda$\,1238,1242 to have a single velocity component centered at $\approx -32$\,km\,s$^{-1}$ relative to $z_{\mathrm{GRB}}$. 
The spectral region around the lines S\,\textsc{vi}\,$\lambda\lambda$\,933,944 are not covered by the UVB arm. O\,\textsc{vi}\,$\lambda\lambda$\,1031,1037 are also located in a region with a poor to no spectral signal. The X-ray derived metal column density used in Sect.~\ref{ssec:cbm} for this burst is from the \textit{Swift}/XRT database\footnote{\url{http://www.swift.ac.uk/xrt_spectra/00637044/}}.

\subsection{GRB\,151021A at $z=2.32975$}

\begin{figure}
	\centering
	\epsfig{file=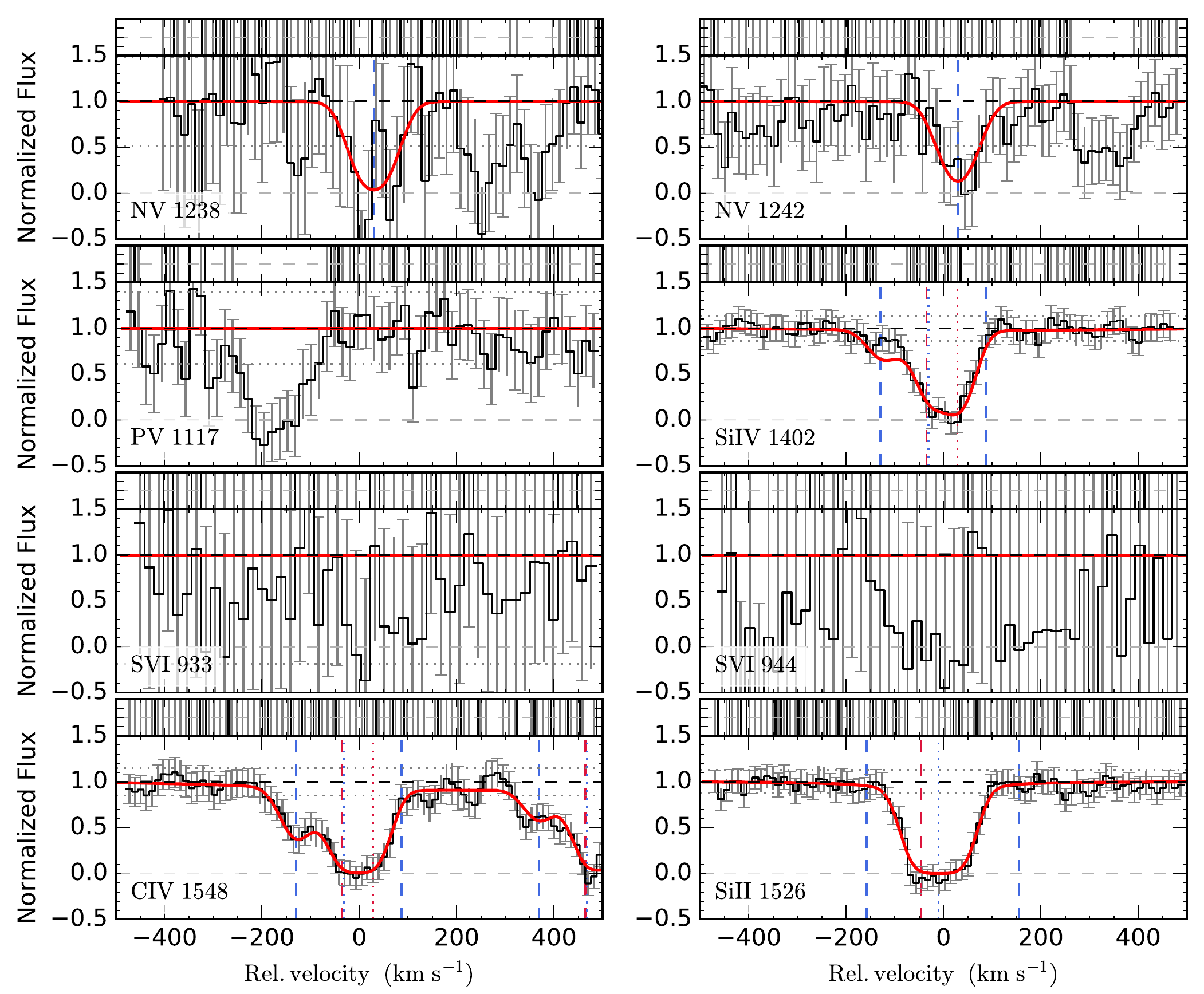,page=1,width=8.9cm}
	\epsfig{file=GRB151021A_hion.pdf,page=2,width=8.9cm}
	\caption{GRB151021A.}
	\label{fig:151021A_hion}
\end{figure}

The spectrum of GRB\,151021A was secured 0.75 hours after trigger using the rapid-response mode with an acquisition magnitude of 18.2 mag. The H\,\textsc{i} column density is adopted from Tanvir et al. (submitted) and was used for the determination of the metallicity. Th\"one et al. (in preparation) measure [Si/H] = $-1.11\pm 0.20$ from a combined fit to  Si\,\textsc{ii}\,$\lambda\lambda$\,1526,1808. We detect high-ionization absorption features from N\,\textsc{v}, C\,\textsc{iv}, and Si\,\textsc{iv} (see Fig.~\ref{fig:151021A_hion}). Strong absorption from Si\,\textsc{iv} and C\,\textsc{iv} are observed and both features are mildly saturated. The five velocity components of both of these line profiles are similar to the low-ionization absorption features, represented in the figure by Si\,\textsc{ii}\,$\lambda$\,1526, and the total line profiles extend from $-200$\,km\,s$^{-1}$ to $+100$\,km\,s$^{-1}$. We report a weak detection of N\,\textsc{v}\,$\lambda\lambda$\,1238,1242 with a single velocity component centered at $\approx 34$\,km\,s$^{-1}$ relative to $z_{\mathrm{GRB}}$. For the remaining high-ionization absorption lines we surveyed we report a non-detection but caution that the spectral region around those lines have poor S/N. The X-ray derived metal column density used in Sect.~\ref{ssec:cbm} for this burst is from the \textit{Swift}/XRT database\footnote{\url{http://www.swift.ac.uk/xrt_spectra/00660671/}}.

\subsection{GRB\,151027B at $z=4.06463$}

\begin{figure}
	\centering
	\epsfig{file=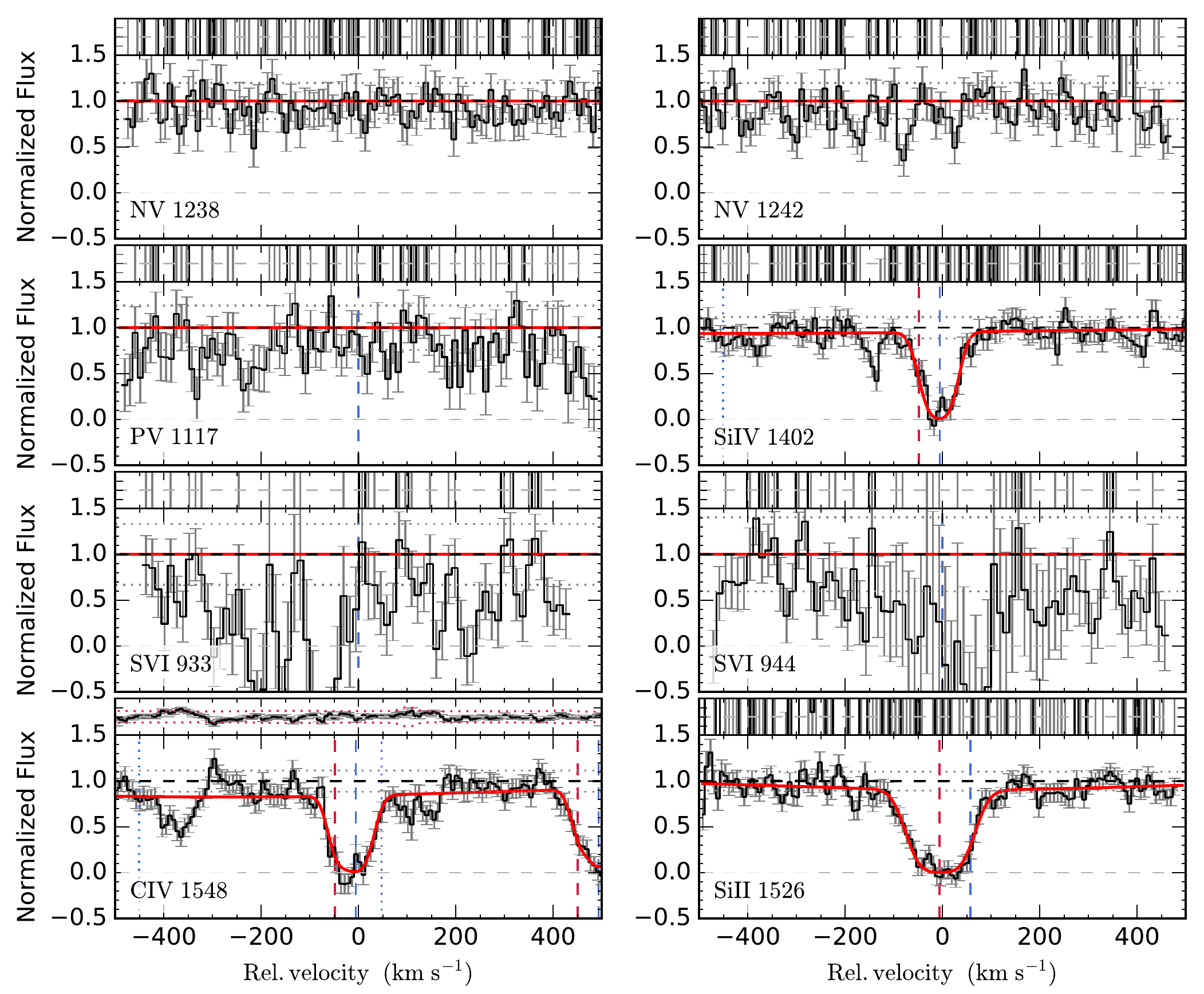,page=1,width=8.9cm}
	\epsfig{file=GRB151027B_hion.pdf,page=2,width=8.9cm}
	\caption{GRB151027B.}
	\label{fig:151027B_hion}
\end{figure}

The spectrum of GRB\,151027B was secured 5 hours after trigger with an acquisition magnitude of 20.5 mag. The H\,\textsc{i} column density is adopted from Tanvir et al. (submitted) and was used for the determination of the metallicity. Th\"one et al. (in preparation) measure [Si/H] = $-1.62\pm 0.24$ from a fit to Si\,\textsc{ii}\,$\lambda$\,1526. We detect high-ionization absorption features from C\,\textsc{iv} and Si\,\textsc{iv} (see Fig.~\ref{fig:151027B_hion}) and measure an upper limit for N\,\textsc{v} of $\log N$(N\,\textsc{v}) < 13.98. We report a non-detections of N\,\textsc{v} and P\,\textsc{v} but caution that the spectral regions covering these two lines have poor S/N. S\,\textsc{vi}, O\,\textsc{vi}, and S\,\textsc{iv}, if present, are all blended. The velocity components of C\,\textsc{iv} and Si\,\textsc{iv} follow the same distribution as the low-ionization absorption line profiles, represented in the figure by Si\,\textsc{ii}\,$\lambda$\,1526. They show two components separated by only $\approx 10$\,km\,s$^{-1}$, but the total line profile covers $\pm 100$\,km\,s$^{-1}$ around $z_{\mathrm{GRB}}$. The X-ray derived metal column density used in Sect.~\ref{ssec:cbm} for this burst is from the \textit{Swift}/XRT database\footnote{\url{http://www.swift.ac.uk/xrt_spectra/00661869/}}.

\subsection{GRB\,160203A at $z=3.51871$}

\begin{figure}
	\centering
	\epsfig{file=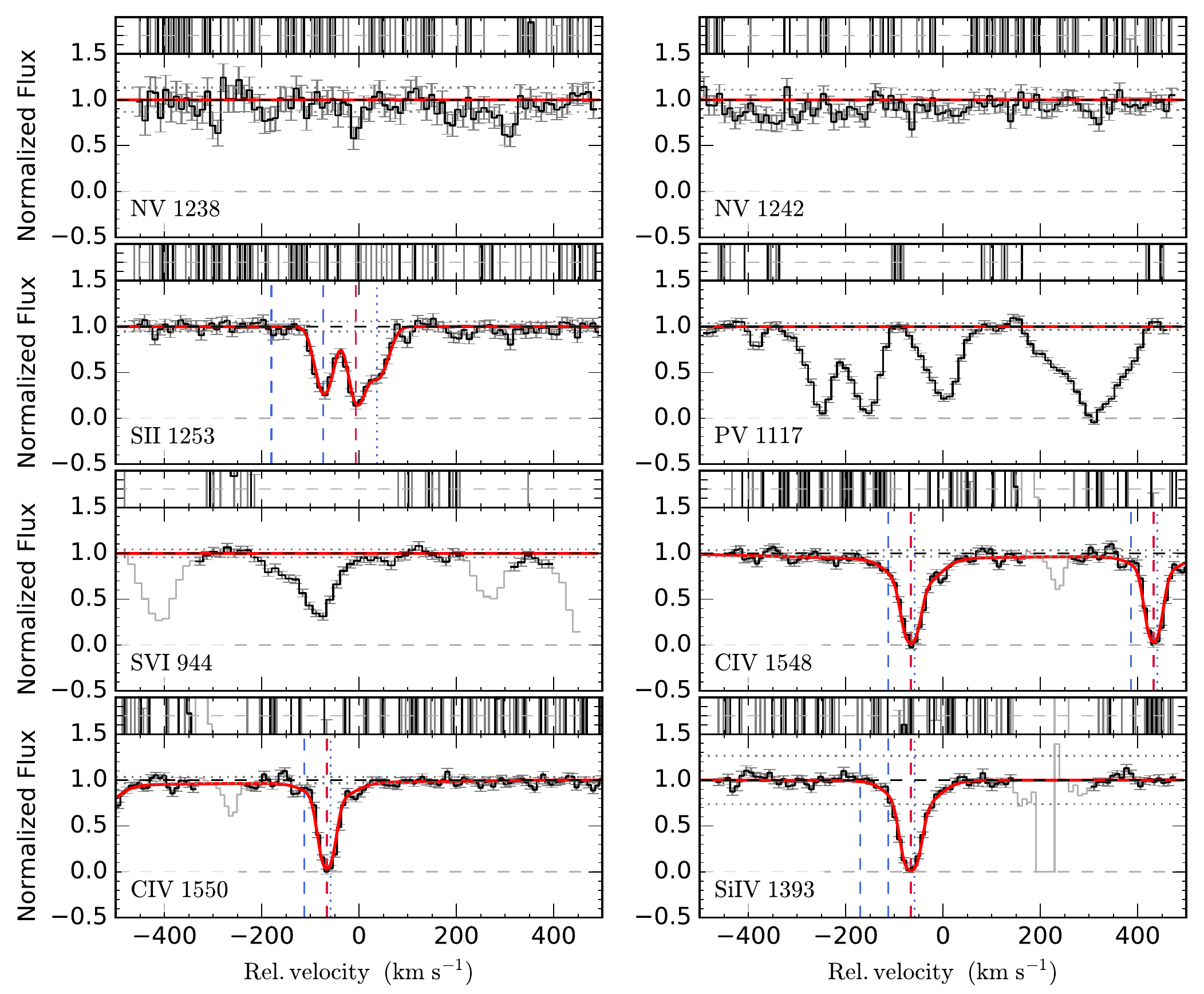,page=1,width=8.9cm}
	\epsfig{file=GRB160203A_hion.pdf,page=2,width=8.9cm}
	\caption{GRB160203A.}
	\label{fig:160203A_hion}
\end{figure}

The spectrum of GRB\,161023A was secured 0.3 hours after trigger using the rapid-response mode with an acquisition magnitude of 18.0 mag. The H\,\textsc{i} column density is adopted from Tanvir et al. (submitted) and was used for the determination of the metallicity. Th\"one et al. (in preparation) measure [S/H] = $-1.26\pm 0.11$ from a combined fit to S\,\textsc{ii}\,$\lambda\lambda\lambda$\,1250,1253,1259. We detect high-ionization absorption features from C\,\textsc{iv}, Si\,\textsc{iv}, S\,\textsc{iv}, and S\,\textsc{vi} (see Fig.~\ref{fig:160203A_hion}) and measure an upper limit for N\,\textsc{v} of $\log N$(N\,\textsc{v}) < 13.58. Strong absorption from Si\,\textsc{iv} and C\,\textsc{iv} are observed and both features are mildly saturated with a single (Si\,\textsc{iv}) or two (C\,\textsc{iv}) velocity components. We report a non-detection of N\,\textsc{v} in this burst, but caution that the spectral region around the absorption feature has poor S/N. The most prominent high-ionization absorption features are centered at $\approx -70$\,km\,s$^{-1}$. There are tentative detections of O\,\textsc{vi} and P\,\textsc{v} but these features are heavily blended with unrelated lines from the Ly$\alpha$ forest. We were thus not able to securily fit an absorption profile to these lines. The X-ray derived metal column density used in Sect.~\ref{ssec:cbm} for this burst is from the \textit{Swift}/XRT database\footnote{\url{http://www.swift.ac.uk/xrt_spectra/00672525/}}.

\subsection{GRB\,161023A at $z=2.71067$}

The spectrum of GRB\,161023A was secured 3 hours after trigger with an acquisition magnitude of 17.5 mag. The H\,\textsc{i} column density is adopted from Tanvir et al. (submitted) and was used for the determination of the metallicity. Th\"one et al. (in preparation) measure [Si/H] = $-1.24\pm 0.09$ from a combined fit to Si\,\textsc{ii}\,$\lambda\lambda\lambda$\,1260,1526,1808. We detect high-ionization absorption features from N\,\textsc{v}, O\,\textsc{vi}, C\,\textsc{iv}, Si\,\textsc{iv}, S\,\textsc{iv}, S\,\textsc{vi}, and P\,\textsc{v} (see Fig.~\ref{fig:161023A_hion}). Strong absorption from Si\,\textsc{iv} and C\,\textsc{iv} are observed and both features are mildly saturated with 2 -- 3 velocity components. A single component at $\approx -120$\,km\,s$^{-1}$ is seen in N\,\textsc{v}, P\,\textsc{v}, S\,\textsc{vi}, O\,\textsc{vi}, S\,\textsc{iv}. This is in contrast with the multiple velocity components seen in the low-ionization absorption features represented by Fe\,\textsc{ii}\,$\lambda$\,1608 in Fig.~\ref{fig:161023A_hion}. We caution that the lines S\,\textsc{vi}\,$\lambda$\,933, O\,\textsc{vi}\,$\lambda\lambda$\,1031,1037, S\,\textsc{iv}\,$\lambda$\,1062, and P\,\textsc{v}\,$\lambda$\,1128 are blended with unrelated absorption features from the Ly$\alpha$ forest. The X-ray derived metal column density used in Sect.~\ref{ssec:cbm} for this burst is from the \textit{Swift}/XRT database\footnote{\url{http://www.swift.ac.uk/xrt_spectra/00020709/}}.


\section{Voigt profile fits - tables \& figures}

\begin{table}
	\caption{Voigt profile fits for GRB\,090809 at $z=2.73706$. }
	\begin{tabular}{ccccc}
		\noalign{\smallskip} \hline \hline \noalign{\smallskip}
		Ion & $v_0$ & $b$ & $\log N$ & $\log N(\mathrm{total})$  \\
		& (km s$^{-1}$)  & (km s$^{-1}$) & ($N$ in cm$^{-2}$) & ($N$ in cm$^{-2}$) \\ 
		\noalign{\smallskip} \hline \noalign{\smallskip}
		N\,\textsc{v} & $\cdots$ & $\cdots$ & $\cdots$ & $<14.43$ \\
		\noalign{\smallskip} \hline \noalign{\smallskip}
	\end{tabular}	
	\label{tab:090809}
\end{table}

\begin{table}
	\caption{Voigt profile fits for GRB\,090926A at $z=2.10694$. }
	\begin{tabular}{ccccc}
		\noalign{\smallskip} \hline \hline \noalign{\smallskip}
		Ion & $v_0$ & $b$ & $\log N$ & $\log N(\mathrm{total})$  \\
		& (km s$^{-1}$)  & (km s$^{-1}$) & ($N$ in cm$^{-2}$) & ($N$ in cm$^{-2}$) \\ 
		\noalign{\smallskip} \hline \noalign{\smallskip}
		N\,\textsc{v} & $4\pm 2$ & $15\pm 3$ & $14.30\pm 0.19$ & $14.30\pm 0.19$ \\
		C\,\textsc{iv} & $-6\pm 5$ & $16\pm 3$ & $15.94\pm 0.78$ & $15.94\pm 0.78$ \\
		& $-76\pm 14$ & $63\pm 13$ & $13.62\pm 0.11$ & \\
		Si\,\textsc{iv} & $-6\pm 5$ & $16\pm 3$ & $15.19\pm 0.62$ & $15.19\pm 0.62$ \\
		& $-76\pm 14$ & $63\pm 13$ & $13.32\pm 0.11$ & \\
		S\,\textsc{iv} & $0\pm 2$ & $15\pm 3$ & $14.79\pm 0.15$ & $14.79\pm 0.15$ \\ 
		\noalign{\smallskip} \hline \noalign{\smallskip}
	\end{tabular}	
	\label{tab:090926a}
\end{table}

\begin{table}
	\caption{Voigt profile fits for GRB\,100425A at $z=1.75640$.}
	\begin{tabular}{ccccc}
		\noalign{\smallskip} \hline \hline \noalign{\smallskip}
		Ion & $v_0$ & $b$ & $\log N$ & $\log N(\mathrm{total})$  \\
		& (km s$^{-1}$)  & (km s$^{-1}$) & ($N$ in cm$^{-2}$) & ($N$ in cm$^{-2}$) \\ 
		\noalign{\smallskip} \hline \noalign{\smallskip}
		N\,\textsc{v} & $\cdots$ & $\cdots$ & $\cdots$ & $<14.63$ \\
		\noalign{\smallskip} \hline \noalign{\smallskip}
	\end{tabular}	
	\label{tab:100425a}
\end{table}

\begin{table}
	\caption{Voigt profile fits for GRB\,111008A at $z=4.99146$.}
	\begin{tabular}{ccccc}
		\noalign{\smallskip} \hline \hline \noalign{\smallskip}
		Ion & $v_0$ & $b$ & $\log N$ & $\log N(\mathrm{total})$  \\
		& (km s$^{-1}$)  & (km s$^{-1}$) & ($N$ in cm$^{-2}$) & ($N$ in cm$^{-2}$) \\ 
		\noalign{\smallskip} \hline \noalign{\smallskip}
		N\,\textsc{v}$^a$ & $8\pm 10$ & $50\pm 15$ & $>14.00$ & $>14.00$ \\
		\noalign{\smallskip} \hline \noalign{\smallskip}
	\end{tabular}	
	\label{tab:111008a}
\end{table}

\begin{table}
	\caption{Voigt profile fits for GRB\,120119A at $z=1.72883$.}
	\begin{tabular}{ccccc}
		\noalign{\smallskip} \hline \hline \noalign{\smallskip}
		Ion & $v_0$ & $b$ & $\log N$ & $\log N(\mathrm{total})$  \\
		& (km s$^{-1}$)  & (km s$^{-1}$) & ($N$ in cm$^{-2}$) & ($N$ in cm$^{-2}$) \\ 
		\noalign{\smallskip} \hline \noalign{\smallskip}
		N\,\textsc{v} & $\cdots$ & $\cdots$ & $\cdots$ & $<15.15$ \\
		\noalign{\smallskip} \hline \noalign{\smallskip}
	\end{tabular}
	\label{tab:120119a}
\end{table}

\begin{table}
	\caption{Voigt profile fits for GRB\,120815A at $z=2.35820$.}
	\begin{tabular}{ccccc}
		\noalign{\smallskip} \hline \hline \noalign{\smallskip}
		Ion & $v_0$ & $b$ & $\log N$ & $\log N(\mathrm{total})$  \\
		& (km s$^{-1}$)  & (km s$^{-1}$) & ($N$ in cm$^{-2}$) & ($N$ in cm$^{-2}$) \\ 
		\noalign{\smallskip} \hline \noalign{\smallskip}
		N\,\textsc{v} & $-6\pm 7$ & $16\pm 8$ & $14.60\pm 0.18$ & $14.60\pm 0.18$ \\
		C\,\textsc{iv} & $0\pm 2$ & $18\pm 3$ & $17.54\pm 0.18$ & $17.55\pm 0.18$ \\
		& $-23\pm 2$ & $38\pm 5$ & $15.46\pm 0.12$ &  \\ 
		Si\,\textsc{iv} & $0\pm 2$ & $18\pm 3$ & $17.40\pm 0.43$ & $17.40\pm 0.43$ \\
		& $-23\pm 2$ & $38\pm 5$ & $14.87\pm 0.37$ &  \\ 
		\noalign{\smallskip} \hline \noalign{\smallskip}
	\end{tabular}	
	\label{tab:120815a}
\end{table}

\begin{table}
	\caption{Voigt profile fits for GRB\,120909A at $z=3.92882$.}
	\begin{tabular}{ccccc}
		\noalign{\smallskip} \hline \hline \noalign{\smallskip}
		Ion & $v_0$ & $b$ & $\log N$ & $\log N(\mathrm{total})$  \\
		& (km s$^{-1}$)  & (km s$^{-1}$) & ($N$ in cm$^{-2}$) & ($N$ in cm$^{-2}$) \\ 
		\noalign{\smallskip} \hline \noalign{\smallskip}
		N\,\textsc{v} & $-14\pm 4$ & $41\pm 6$ & $14.75\pm 0.11$ & $14.75\pm 0.11$ \\
		\noalign{\smallskip} \hline \noalign{\smallskip}
	\end{tabular}	
	\label{tab:120909a}
\end{table}

\begin{table}
	\caption{Voigt profile fits for GRB\,121024A at $z=2.30244$.}
	\begin{tabular}{ccccc}
		\noalign{\smallskip} \hline \hline \noalign{\smallskip}
		Ion & $v_0$ & $b$ & $\log N$ & $\log N(\mathrm{total})$  \\
		& (km s$^{-1}$)  & (km s$^{-1}$) & ($N$ in cm$^{-2}$) & ($N$ in cm$^{-2}$) \\ 
		\noalign{\smallskip} \hline \noalign{\smallskip}
		N\,\textsc{v} & $\cdots$ & $\cdots$ & $\cdots$ & $<14.35$ \\
		\noalign{\smallskip} \hline \noalign{\smallskip}
	\end{tabular}	
	\label{tab:121024a}
\end{table}

\begin{table}
	\caption{Voigt profile fits for GRB\,130408A at $z=3.75792$.}
	\begin{tabular}{ccccc}
		\noalign{\smallskip} \hline \hline \noalign{\smallskip}
		Ion & $v_0$ & $b$ & $\log N$ & $\log N(\mathrm{total})$  \\
		& (km s$^{-1}$)  & (km s$^{-1}$) & ($N$ in cm$^{-2}$) & ($N$ in cm$^{-2}$) \\ 
		\noalign{\smallskip} \hline \noalign{\smallskip}
		N\,\textsc{v} & $-16\pm 2$ & $31\pm 4$ & $14.44\pm 0.07$ & $14.44\pm 0.07$ \\
		C\,\textsc{iv} & $65\pm 3$ & $16\pm 2$ & $13.71\pm 0.04$ & $15.12\pm 0.05$ \\
		& $0\pm 2$ & $13\pm 3$ & $10.83\pm 0.01$ &  \\ 
		& $-17\pm 2$ & $28\pm 14$ & $15.11\pm 0.05$ &  \\ 
		Si\,\textsc{iv} & $65\pm 3$ & $16\pm 2$ & $13.12\pm 0.02$ & $14.42\pm 0.05$ \\
		& $0\pm 2$ & $13\pm 3$ & $13.60\pm 0.02$ &  \\ 
		& $-17\pm 2$ & $28\pm 14$ & $14.32\pm 0.04$ &  \\ 
		S\,\textsc{vi} & $65\pm 3$ & $16\pm 2$ & $13.50\pm 0.02$ & $14.37\pm 0.05$ \\
		& $0\pm 2$ & $13\pm 3$ & $13.75\pm 0.02$ &  \\ 
		& $-17\pm 2$ & $28\pm 14$ & $14.17\pm 0.04$ &  \\ 
		\noalign{\smallskip} \hline \noalign{\smallskip}
	\end{tabular}	
	\label{tab:130408a}
\end{table}

\begin{table}
	\caption{Voigt profile fits for GRB\,130606A at $z=5.91278$. }
	\begin{tabular}{ccccc}
		\noalign{\smallskip} \hline \hline \noalign{\smallskip}
		Ion & $v_0$ & $b$ & $\log N$ & $\log N(\mathrm{total})$  \\
		& (km s$^{-1}$)  & (km s$^{-1}$) & ($N$ in cm$^{-2}$) & ($N$ in cm$^{-2}$) \\ 
		\noalign{\smallskip} \hline \noalign{\smallskip}
		N\,\textsc{v} & $64\pm 1$ & $13\pm 2$ & $13.78\pm 0.03$ & $14.50\pm 0.04$ \\
		& $-11\pm 4$ & $57\pm 11$ & $13.92\pm 0.06$ & \\
		& $-83\pm 1$ & $19\pm 1$ & $14.16\pm 0.03$ & \\
		& $-174\pm 4$ & $28\pm 6$ & $13.27\pm 0.07$ & \\
		\noalign{\smallskip} \hline \noalign{\smallskip}
	\end{tabular}	
	\label{tab:130606a}
\end{table}

\begin{table}
	\caption{Voigt profile fits for GRB\,141028A at $z=2.33327$.}
	\begin{tabular}{ccccc}
		\noalign{\smallskip} \hline \hline \noalign{\smallskip}
		Ion & $v_0$ & $b$ & $\log N$ & $\log N(\mathrm{total})$  \\
		& (km s$^{-1}$)  & (km s$^{-1}$) & ($N$ in cm$^{-2}$) & ($N$ in cm$^{-2}$) \\ 
		\noalign{\smallskip} \hline \noalign{\smallskip}
		N\,\textsc{v} & $-104\pm 20$ & $10\pm 4$ & $14.28\pm 0.10$ & $14.28\pm 0.10$ \\
		C\,\textsc{iv} & $96\pm 7$ & $35\pm 2$ & $14.11\pm 0.02$ & $14.96\pm 0.05$ \\
		& $-24\pm 3$ & $66\pm 3$ & $14.17\pm 0.02$ &  \\ 
		& $-88\pm 11$ & $25\pm 14$ & $14.71\pm 0.03$ &  \\ 
		Si\,\textsc{iv} & $107\pm 23$ & $49\pm 3$ & $13.47\pm 0.04$ & $14.94\pm 0.07$ \\
		& $-39\pm 10$ & $67\pm 7$ & $13.69\pm 0.04$ & \\
		& $-89\pm 11$ & $12\pm 4$ & $14.87\pm 0.06$ & \\
		P\,\textsc{v} & $-67\pm 21$ & $28\pm 12$ & $14.86\pm 0.04$ & $14.86\pm 0.04$ \\
		\noalign{\smallskip} \hline \noalign{\smallskip}
	\end{tabular}	
	\label{tab:141028a}
\end{table}

\begin{table}
	\caption{Voigt profile fits for GRB\,141109A at $z=2.99438$. }
	\begin{tabular}{ccccc}
		\noalign{\smallskip} \hline \hline \noalign{\smallskip}
		Ion & $v_0$ & $b$ & $\log N$ & $\log N(\mathrm{total})$  \\
		& (km s$^{-1}$)  & (km s$^{-1}$) & ($N$ in cm$^{-2}$) & ($N$ in cm$^{-2}$) \\ 
		\noalign{\smallskip} \hline \noalign{\smallskip}
		N\,\textsc{v}$^a$ & $13\pm 15$ & $9\pm 3$ & $\gtrsim14.85$ & $\gtrsim14.85$  \\
		\noalign{\smallskip} \hline \noalign{\smallskip}
	\end{tabular}	
	\label{tab:141109a}
\end{table}

\begin{table}
	\caption{Voigt profile fits for GRB\,150403A at $z=2.05707$.}
	\begin{tabular}{ccccc}
		\noalign{\smallskip} \hline \hline \noalign{\smallskip}
		Ion & $v_0$ & $b$ & $\log N$ & $\log N(\mathrm{total})$  \\
		& (km s$^{-1}$)  & (km s$^{-1}$) & ($N$ in cm$^{-2}$) & ($N$ in cm$^{-2}$) \\ 
		\noalign{\smallskip} \hline \noalign{\smallskip}
		N\,\textsc{v} & $-32\pm 3$ & $48\pm 11$ & $14.73\pm 0.14$ & $14.73\pm 0.14$ \\
		C\,\textsc{iv} & $44\pm 2$ & $99\pm 3$ & $14.84\pm 0.07$ & $15.75\pm 0.10$ \\
		& $22\pm 2$ & $69\pm 3$ & $15.16\pm 0.07$ &  \\ 
		& $-50\pm 4$ & $5\pm 3$ & $14.24\pm 0.02$ &  \\ 
		& $-196\pm 5$ & $22\pm 5$ & $15.52\pm 0.04$ &  \\ 
		Si\,\textsc{iv} & $120\pm 4$ & $58\pm 7$ & $13.93\pm 0.03$ & $14.58\pm 0.08$ \\
		& $23\pm 2$ & $55\pm 7$ & $13.89\pm 0.05$ &  \\ 
		& $-10\pm 4$ & $4\pm 2$ & $12.64\pm 0.03$ &  \\ 
		& $-195\pm 5$ & $42\pm 5$ & $14.33\pm 0.04$ &  \\ 
		\noalign{\smallskip} \hline \noalign{\smallskip}
	\end{tabular}	
	\label{tab:150403a}
\end{table}

\begin{table}
	\caption{Voigt profile fits for GRB\,151021A at $z=2.32975$.}
	\begin{tabular}{ccccc}
		\noalign{\smallskip} \hline \hline \noalign{\smallskip}
		Ion & $v_0$ & $b$ & $\log N$ & $\log N(\mathrm{total})$  \\
		& (km s$^{-1}$)  & (km s$^{-1}$) & ($N$ in cm$^{-2}$) & ($N$ in cm$^{-2}$) \\ 
		\noalign{\smallskip} \hline \noalign{\smallskip}
		N\,\textsc{v} & $34\pm 5$ & $31\pm 19$ & $14.80\pm 0.46$ & $14.80\pm 0.46$ \\
		C\,\textsc{iv} & $47\pm 2$ & $15\pm 3$ & $12.21\pm 0.07$ & $18.26\pm 0.17$ \\
		& $0\pm 2$ & $19\pm 3$ & $18.26\pm 0.10$ &  \\ 
		& $-52\pm 4$ & $18\pm 3$ & $11.25\pm 0.02$ &  \\ 
		& $-121\pm 5$ & $35\pm 5$ & $13.90\pm 0.04$ &  \\ 
		& $-160\pm 7$ & $2\pm 3$ & $14.70\pm 0.06$ &  \\ 
		Si\,\textsc{iv} & $47\pm 4$ & $15\pm 7$ & $13.88\pm 0.03$ & $15.32\pm 0.07$ \\
		& $0\pm 2$ & $19\pm 3$ & $15.28\pm 0.05$ &  \\ 
		& $-52\pm 4$ & $18\pm 3$ & $13.55\pm 0.03$ &  \\ 
		& $-121\pm 5$ & $35\pm 5$ & $13.57\pm 0.04$ &  \\ 
		& $-160\pm 7$ & $2\pm 3$ & $12.27\pm 0.02$ &  \\ 
		\noalign{\smallskip} \hline \noalign{\smallskip}
	\end{tabular}	
	\label{tab:151021a}
\end{table}

\begin{table}
	\caption{Voigt profile fits for GRB\,151027B at $z=4.06463$.}
	\begin{tabular}{ccccc}
		\noalign{\smallskip} \hline \hline \noalign{\smallskip}
		Ion & $v_0$ & $b$ & $\log N$ & $\log N(\mathrm{total})$  \\
		& (km s$^{-1}$)  & (km s$^{-1}$) & ($N$ in cm$^{-2}$) & ($N$ in cm$^{-2}$) \\ 
		\noalign{\smallskip} \hline \noalign{\smallskip}
		N\,\textsc{v} & $\cdots$ & $\cdots$ & $\cdots$ & $<13.98$ \\
		C\,\textsc{iv} & $0\pm 2$ & $55\pm 3$ & $13.53\pm 0.04$ & $14.82\pm 0.09$ \\
		& $-9\pm 2$ & $28\pm 3$ & $14.80\pm 0.08$ &  \\ 
		Si\,\textsc{iv} & $0\pm 2$ & $55\pm 7$ & $13.64\pm 0.04$ & $14.38\pm 0.10$ \\
		& $-9\pm 2$ & $28\pm 3$ & $14.29\pm 0.08$ &  \\ 
		\noalign{\smallskip} \hline \noalign{\smallskip}
	\end{tabular}	
	\label{tab:151027b}
\end{table}

\begin{table}
	\caption{Voigt profile fits for GRB\,160203A at $z=3.51871$.}
	\begin{tabular}{ccccc}
		\noalign{\smallskip} \hline \hline \noalign{\smallskip}
		Ion & $v_0$ & $b$ & $\log N$ & $\log N(\mathrm{total})$  \\
		& (km s$^{-1}$)  & (km s$^{-1}$) & ($N$ in cm$^{-2}$) & ($N$ in cm$^{-2}$) \\ 
		\noalign{\smallskip} \hline \noalign{\smallskip}
		N\,\textsc{v} & $\cdots$ & $\cdots$ & $\cdots$ & $<13.58$ \\
		C\,\textsc{iv} & $-64\pm 2$ & $17\pm 3$ & $14.54\pm 0.07$ & $14.56\pm 0.07$ \\
		& $-4\pm 2$ & $15\pm 3$ & $13.22\pm 0.04$ &  \\ 
		Si\,\textsc{iv} & $-117\pm 4$ & $26\pm 7$ & $14.47\pm 0.18$ & $14.50\pm 0.17$ \\
		S\,\textsc{iv} & $-68\pm 4$ & $39\pm 2$ & $14.46\pm 0.07$ & $15.79\pm 0.14$ \\ 
		& $-173\pm 34$ & $28\pm 20$ & $14.77\pm 0.08$ &  \\ 
		& $-230\pm 34$ & $42\pm 20$ & $15.72\pm 0.13$ &  \\ 
		S\,\textsc{vi} & $-89\pm 6$ & $49\pm 8$ & $14.22\pm 0.06$ & $14.22\pm 0.06$ \\
		\noalign{\smallskip} \hline \noalign{\smallskip}
	\end{tabular}	
	\label{tab:160203a}
\end{table}


\clearpage


\bsp    
\label{lastpage}
\end{document}